\documentclass[twocolumn]{emulateapj}

\usepackage{graphicx}
\usepackage{epsfig,lscape}

\newcommand\hi{H$\,${\small I}}
\newcommand\hii{H$\,${\small II}}
\newcommand\halpha{H$\alpha$}
\newcommand\kms{$\rm km\;s^{-1}$}
\newcommand\dg{$^{\circ}$}
\newcommand\tHI{$t_{{\rm HI}\mapsto 24\,\mu{\rm m}}$}
\newcommand\tCO{$t_{{\rm CO}\mapsto {\rm H}\alpha}$}
\newcommand\mum{$\mu$m}
\newcommand\der{{\rm d}}

\slugcomment{Accepted for publication in the AJ special THINGS issue. For  
a high-resolution version visit: http://www.mpia.de/THINGS/Publications.html}

\shorttitle{Timescales for Star Formation in Spiral Galaxies}
\shortauthors{D.Tamburro, et al.}

\begin{document}

\title{GEOMETRICALLY DERIVED TIMESCALES FOR STAR FORMATION IN SPIRAL GALAXIES}

\author{D. Tamburro, H.-W. Rix and F. Walter}
\affil{Max-Planck-Institut f\"ur Astronomie, K\"onigstuhl 17,
 D-69117 Heidelberg, Germany}
\email{tamburro@mpia.de, rix@mpia.de, walter@mpia.de}

\author{E. Brinks}
\affil{Centre for Astrophysics Research, University of Hertfordshire, College
  Lane, Hatfield  AL10 9AB, United Kingdom}
\email{e.brinks@herts.ac.uk}

\author{W.J.G. de Blok} \affil{Department of Astronomy, University of Cape
  Town, Private Bag X3, Rondebosch 7701, South Africa}
\email{edeblok@circinus.ast.uct.ac.za}

\author{R.C. Kennicutt}
\affil{Institute of Astronomy, University of Cambridge,
  Madingley Road, Cambridge CB3 0HA, United Kingdom}
\email{robk@ast.cam.ac.uk}

\and

\author{M.-M. Mac Low\altaffilmark{1}}
\affil{Department of Astrophysics, American Museum of Natural History,
79th Street and Central Park West, New York, NY 10024-5192, USA}
\email{mordecai@amnh.org}
\altaffiltext{1}{also Max-Planck-Institut f\"ur Astronomie, and
Institut f\"ur Theoretische Astrophysik, Zentrum f\"ur Astronomie
der Universit\"at Heidelberg}

\begin{abstract}
  We estimate a characteristic timescale for star formation in the spiral arms
  of disk galaxies, going from atomic hydrogen (\hi) to dust-enshrouded
  massive stars.  Drawing on high-resolution \hi\ data from The \hi\ Nearby
  Galaxy Survey and 24~$\mu$m images from the {\em Spitzer} Infrared Nearby
  Galaxies Survey we measure the average angular offset between the \hi\ and
  24$\mu$m emissivity peaks as a function of radius, for a sample of 14 nearby
  disk galaxies.  We model these offsets assuming an instantaneous kinematic
  pattern speed, $\Omega_p$, and a timescale, \tHI , for the characteristic
  time span between the dense \hi\ phase and the formation of massive stars
  that heat the surrounding dust. Fitting for $\Omega_p$ and \tHI , we find
  that the radial dependence of the observed angular offset (of the \hi\ and
  24~$\mu$m emission) is consistent with this simple prescription; the
  resulting corotation radii of the spiral patterns are typically $R_{\rm
    cor}\simeq 2.7 R_{s}$, consistent with independent estimates.  The
  resulting values of \tHI\ for the sample are in the range 1--4 Myr. We have
  explored the possible impact of non-circular gas motions on the estimate of
  \tHI\ and have found it to be substantially less than a factor of 2. This
  implies that a short timescale for the most intense phase of the ensuing
  star formation in spiral arms, and implies a considerable fraction of
  molecular clouds exist only for a few Myr before forming stars.  However,
  our analysis does not preclude that some molecular clouds persist
  considerably longer. If much of the star formation in spiral arms occurs
  within this short interval \tHI, then star formation must be inefficient, in
  order to avoid the short-term depletion of the gas reservoir.
\end{abstract}


\maketitle

\section{INTRODUCTION}

\citet*[][hereafter R69]{roberts69} was the first to develop the scenario of
spiral-arm-driven star formation in galaxy disks. In this picture a spiral
density wave induces gravitational compression and shocks in the neutral
hydrogen gas, which in turn leads to the collapse of (molecular) gas clouds
that results in star formation. This work already pointed out the basic
consequences for the relative geometry of the dense cold gas
reservoir\footnote{This paper pre-dates observational studies of molecular gas
  in galaxies.} and the emergent young stars: when viewed from a reference
frame that corotates with the density wave, the densest part of the atomic
hydrogen (\hi) lies at the shock (or just upstream from it), while the young
stars lie downstream from the density wave. Using \hi\ and \halpha\ as the
tracers of the cold gas and of the young stars, respectively, R69 found a
qualitative support in the data available at the time. In this picture, the
characteristic timescale for this sequence of events is reflected in the
typical angular offset, at a given radius, between tracers of the different
stages of spiral-arm-driven star formation.

While this qualitative picture has had continued popularity, quantitative
tests of the importance of spiral density waves as star-formation trigger
\citep*{linshu64}\ and of the timescales for the ensuing star formation have
proven complicated. First, it has become increasingly clear that even in
galaxies with grand-design spiral arms, about half of the star formation
occurs in locations outside the spiral arms \citep*{elmegreen86}. Second,
stars formed from molecular clouds (not directly from \hi) and very young star
clusters are dust enshrouded at first.  Moreover, the actual physical
mechanism that appears to control the rate and overall location of star
formation in galaxies is the gravitational instability of the gas and existing
stars \citep*[e.g.][]{li05}.  Stars only form above a critical density
\citep*{martin01} which is consistent with the predicted \citet*{toomre64}\ 
criterion for gravitational instability as generalized by \citet*{rafikov01}.
Although, in galaxies with prominent spiral structure local gas condensations
are governed by magneto-rotational instabilities---spiral arms are regions of
low shear where the transfer of angular momentum is carried out by magnetic
fields \citep{kim02,kim06}. Obtaining high-resolution, sensitive maps of all
phases in this scenario (\hi, molecular gas, dust-enshrouded young stars,
unobscured young stars) has proven technically challenging.

If the star formation originates from direct collapse of gravitationally
unstable gas, and if the rotation curve and approximate pattern speed of the
spiral arms are known, the geometric test suggested by R69 provides a
timescale for the end-to-end (from \hi\ to young stars) process of star
formation. Of course, there are other ways of estimating the timescales that
characterize the evolutionary sequence of the interstellar medium (ISM), based
on other physical arguments. However, other lines of reasoning have led to
quite a wide range of varying lifetime estimates as discussed below.

Offsets between components such as CO and \halpha\ emission in the disks of
spiral galaxies have indeed been observed
\citep*{vogel88,garcia93,rand90,scoville01}. \citet*{mouschovias06} remarked
that the angular separation between the dust lanes and the peaks of \halpha\ 
emission found for nearby spiral galaxies \citep*[e.g. observed
by][]{roberts69,rots75}\ implied timescales of the order of 10~Myr. More
recently \citet*{egusa04}, using the angular offset between CO and \halpha\ in
nearby galaxies, derived \tCO~$\simeq4.8$~Myr.

Observationally, the \hi\ surface density is found to correlate well with
sites of star formation and emission from molecular clouds
\citep*{wong02,kennicutt98}. The conversion timescale of \hi$\mapsto$H$_2$ is
a key issue since it determines how well the peaks of \hi\ emission can be
considered as potential early stages of star formation. H$_2$ molecules only
form on dust grain surfaces in dusty regions that shield the molecules from
ionizing UV photons. Their formation facilitates the subsequent building up of
more complex molecules \citep*[e.g.][]{williams05}.  Within shielded clouds
the conversion timescale \hi$\mapsto$H$_2$ is given by $\tau_{{\rm H}_2}\sim
10^9/n_0\;$yr, where $n_0$ is the proton density in cm$^{-3}$
\citep*{hollenbach71,jura75,goldsmith2005,goldsmith2006}.  Given the inverse
proportionality with $n_0$, the conversion timescale can vary from the edge of
a molecular cloud ($\tau\simeq 4\times10^6\;$yr, $n_0\sim10^3$) to the central
region ($\tau\sim10^5\;$yr, $n_0\sim10^4$) where the density is higher. Local
turbulent compression can further enhance the local density, and thus decrease
the conversion timescale \citep*{glover07}. Thus, even short cloud-formation
timescales remain consistent with the \hi$\mapsto$H$_2$ conversion timescale.

The subsequent evolution \citep*[see e.g. ][ for a review]{beuther06}\ 
involves the formation of cloud cores (initially starless) and then star
cluster formation through accretion onto protostars, which finally become main
sequence stars.  High-mass stars evolve more rapidly than low-mass stars.
Stars with $M\geq5 M_\odot$ reach the main sequence quickly, in less than 1
Myr \citep*{hillenbrand93,palla99}, while they are still deeply embedded and
actively accreting. The O and B stars begin to produce an intense UV flux that
photoionizes the surrounding dusty environment within a few Myr, and
subsequently become optically visible \citep*{thronson86}.

A different scenario is suggested by \citet*{allen02}, in which young stars in
the disks of galaxies produce \hi\ from their parent H$_2$ clouds by
photodissociation.  According to this scenario, the \hi\ should not be seen
furthest upstream in the spiral arm, but rather between the CO and UV/\halpha\ 
regions.  \citet*{allen86}\ indeed report observation of \hi\ downstream of
dust lanes in M83.

Several lines of reasoning, however, point toward longer star-formation
timescales and molecular cloud lifetimes, much greater than 10~Myr.
\citet*{krumholz05}\ conclude that the star-formation rate in the solar
neighborhood is low. In fact, they point out that the star-formation rate in
the solar neighborhood is $\sim100$ times smaller than the ratio of the masses
of nearby molecular clouds to their free-fall time $M_{\rm MC}/\tau_{\rm ff}$,
which also indicates the rate of compression of molecular clouds. Individual
dense molecular clouds have been argued to stay in a fully molecular state for
about 10-15~Myr before their collapse \citep*{tassis04}, and to transform
about 30\% of their mass into stars in $\geq7\:\tau_{\rm ff}$ \citep*[$\sim
10^6\;$yr e.g.  considering the mass of the Orion Nebula Cluster,
ONC][]{tan06}. Large molecular clouds have been calculated to survive 20 to
30~Myr before being destroyed by the stellar feedback by
\citet*{KrumholzMatzner06}.  Based on observations, \citet*{palla99,palla00}\ 
argue that the star formation rate in the ONC was low 10$^7\;$yr ago, and that
it increased only recently.  \citet*{blitz06}, using a statistical comparison
of cluster ages in the Large Magellanic Cloud (LMC) to the presence of CO,
found that the lifetime for giant molecular clouds is 20--30~Myr.

However, other studies conclude that the timescales for star-formation are
rather short.  \citet*{hartmann03}\ pointed out that the Palla--Stahler model
is not consistent with observations since most of the molecular clouds in the
ONC are forming stars at the same high rate. The stellar age or the age spread
in young open stellar clusters is not necessarily a useful constraint on the
star-formation timescale: the age spread, for example, may result from
independent and non-simultaneous bursts of star-formation
\citep*{elmegreen2000}.  \citet*{ballesteros06}\ pointed out that the
molecular cloud lifetime must be shorter than the value of $\tau_{\rm
  MC}\simeq10\;$Myr suggested by \citet*{mouschovias06}.  Also subsequent star
formation must proceed very quickly, within a few Myr
\citep*{vazquez05,HBB01}.  \citet*{prescott07}\ found strong association
between 24$\;\mu$m sources and optical \hii\ regions in nearby spiral
galaxies. This provides constraints on the lifetimes of star-forming clouds:
the break out time of the clouds and their parent clouds is less or at most of
the same order as the lifetime of the \hii\ regions, therefore a few Myr.
Dust and gas clouds must dissipate on a timescale no longer than 5--10 Myr.

In conclusion, all the previous studies listed aim to estimate the lifetimes
of molecular clouds or the timescale separation between the compression of
neutral gas and newly formed stars. Most of these studies are based on
observations of star-forming regions both in the Milky Way and in external
galaxies, and in all cases the derived timescales lie in a range between a few
Myr and several tens of Myr.

In this paper, we examine a new method (\S~\ref{sec:method}) for estimating
the timescale to proceed from \hi\ compression to star formation in nearby
spiral galaxies.  We compare {\em Spitzer Space Telescope}/MIPS $24\:\mu$m
data from the {\em Spitzer} Near Infrared Galaxies Survey (SINGS;
\citealt*{kennicutt03}) to 21~cm maps from the \hi\ Nearby Galaxy Survey
(THINGS; \citealt*{walter07}). The proximity of our targets allows for high
spatial resolution. In \S~\ref{sec:data}\ we give a description of the data.
The MIPS bands (24, 70 and $160\;\mu$m) are tracers of warm dust heated by UV
and are therefore good indicators of recent star-formation activity
\citep*[see for example][]{dale05}. We used the band with the best resolution,
$24\;\mu$m, which has been recognized as the best of the {\em Spitze}r bands
for tracing star formation \citep*{calzetti05,calzetti07,prescott07}; the
$8\;\mu$m {\em Spitzer}/IRAC band has even higher resolution but is
contaminated by PAH features that undergo strong depletion in the presence of
intense UV radiation \citep*{dwek05,smith07}. In \S~\ref{sec:analysis}\ we
describe how we use azimuthal cross-correlation to compare the \hi\ and
$24\;\mu$m images and derive the angular offset of the spiral pattern. This
algorithmic approach minimizes possible biases introduced by subjective
assessments. We describe our results in \S~\ref{sec:results}\ where we derive
\tHI\ for our selection of objects. Finally, we discuss the implications of
our results in \S~\ref{sec:discussion}\ and draw conclusions in
\S~\ref{sec:conclusions}.

\section{METHODOLOGY} \label{sec:method}

The main goal of this paper is to estimate geometrically the timescales for
spiral-arm-driven star formation using a simple kinematic model, examining the
R69 arguments in light of state-of-the-art data.  Specifically, we set out to
determine the relative geometry of two tracers for different stages of
star-formation sequence in a sample of nearby galaxies, drawing on the SINGS
and THINGS data sets (see \S~\ref{sec:data}): the 24~\mum\ and the \hi\ 
emission.

While the angular offset between these two tracers is an empirical
model-independent measurement, a conversion into a star-formation timescale
assumes (a) that peaks of the \hi\ trace material that is forming molecular
clouds, and (b) that the peaks of the 24~\mum\ emission trace the very young,
still dust-enshrouded star clusters, where their UV emission is absorbed and
re-radiated into the mid- to far-infrared wavelength range ($\sim$5~\mum\ to
$\sim$500\mum).  The choice of these particular tracers was motivated by the
fact that they should tightly bracket the conversion process of molecular gas
into young massive stars, and by the availability of high-quality data from
the SINGS and THINGS surveys. Note that a number of imaging studies in the
near-IR have shown \citep*[e.g.  ][]{rix95}\ that the large majority of
luminous disk galaxies have a coherent, dynamically relevant spiral arm
density perturbation.  Therefore, this overall line of reasoning can sensibly
be applied to a sample of disk galaxies.

\begin{figure}
\epsscale{1.0}
\plotone{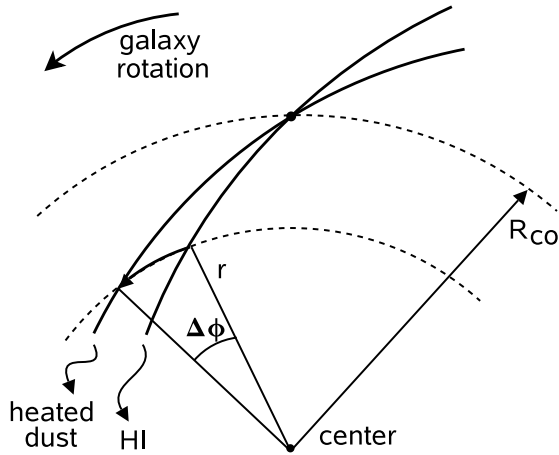}
     \caption{Schematic geometry adopted to derive the azimuthal phase
       difference $(\phi_{24\,\mu{\rm m}}-\phi_{\rm
         HI})(r)\equiv\Delta\phi(r)$ between the \hi\ and the $24\,\mu{\rm m}$
       emission, with $\phi$ increasing in the direction of rotation. The
       sketch shows part of a face-on galaxy rotating anti-clockwise, with the
       center as indicated. The solid curved lines represent the two
       components within one spiral arm, namely the \hi\ and the heated dust.
       The angular separation between the two components is exaggerated for
       clarity. We measured the deprojected phase difference $\Delta\phi(r)$
       at a given radius. Inside corotation, $R_{\rm cor}$, the material is
       rotating faster than the pattern speed and the $24\,\mu{\rm m}$
       emission lies ahead of the \hi\ ($\phi_{\rm HI}<\phi_{24\,\mu{\rm
           m}}$). At corotation the two patterns coincide, and outside $R_{\rm
         cor}$ the picture is reversed since the pattern speed exceeds the
       rotation of the galaxy.}\label{fig:cartoon}
\end{figure}

We consider a radius in the galaxy disk where the spiral pattern can be
described by a kinematic pattern speed, $\Omega_p$, and the local circular
velocity $v_{c}(r)\equiv \Omega (r) \times r$. Then two events separated by a
time \tHI will have a phase offset of
\begin{equation}\label{eq:offset}
\Delta \phi (r) = (\Omega(r)-\Omega_p)\; \mbox{\tHI},
\end{equation}
where \tHI\ denotes the time difference between two particular phases that we
will study here. If the spiral pattern of a galaxy indeed has a characteristic
kinematic pattern speed, the angular offset between any set of tracers is
expected to vary as a function of radius in a characteristic way.  Considering
the chronological sequence, defining the angular phase difference
$\Delta\phi\equiv\phi_{24\,\mu{\rm m}}-\phi_{\rm HI}$ and adopting the
convention that $\phi$ increases in the direction of rotation, we expect the
qualitative radial dependence plotted in Figure~\ref{fig:cartoon}:
$\Delta\phi>0$ where the galaxy rotates faster than the pattern speed,
otherwise $\Delta\phi<0$. Where $\Omega(R_{\rm cor})=\Omega_p$, at the
so-called corotation radius, we expect the sign of $\Delta\phi$ to change.

In practice, the gaseous and stellar distribution is much more complex than in
the qualitative example of Figure~\ref{fig:cartoon}, since the whole spiral
network, even for galaxies where the spiral arms are well defined such as in
grand-design galaxies, typically exhibits a full wealth of smaller scale
sub-structures both in the arms and in the inter-arm regions. The optimal
method to measure the angular offset between the two observed patterns is
therefore through cross-correlation (\S~\ref{sec:analysis}).  We treat the
timescale \tHI\ and the present-day pattern speed $\Omega_p$ as global
constants for each galaxy, although these two parameters might, in principle,
vary as function of galactocentric radius.  Note that we need not to rely on
the assumption that the spiral structure is quasi-stationary over extended
periods, $t\geq t_{\rm dyn}$. Even if spiral arms are quite dynamic,
continuously forming and breaking, and with a pattern speed varying with
radius, our analysis will hold approximately.

\section{DATA}\label{sec:data}

\begin{figure*}
\epsscale{1.0}
\plottwo{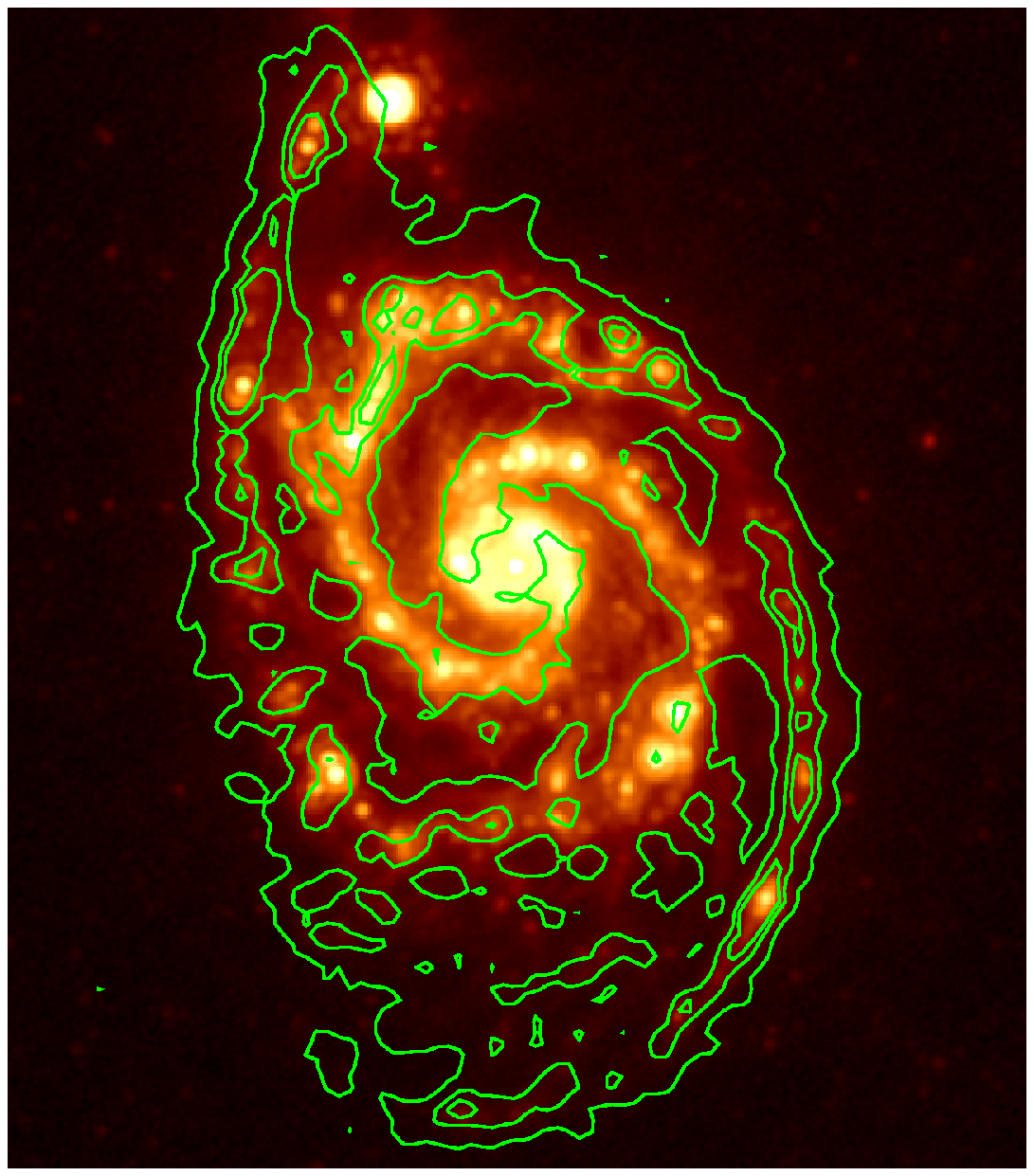}{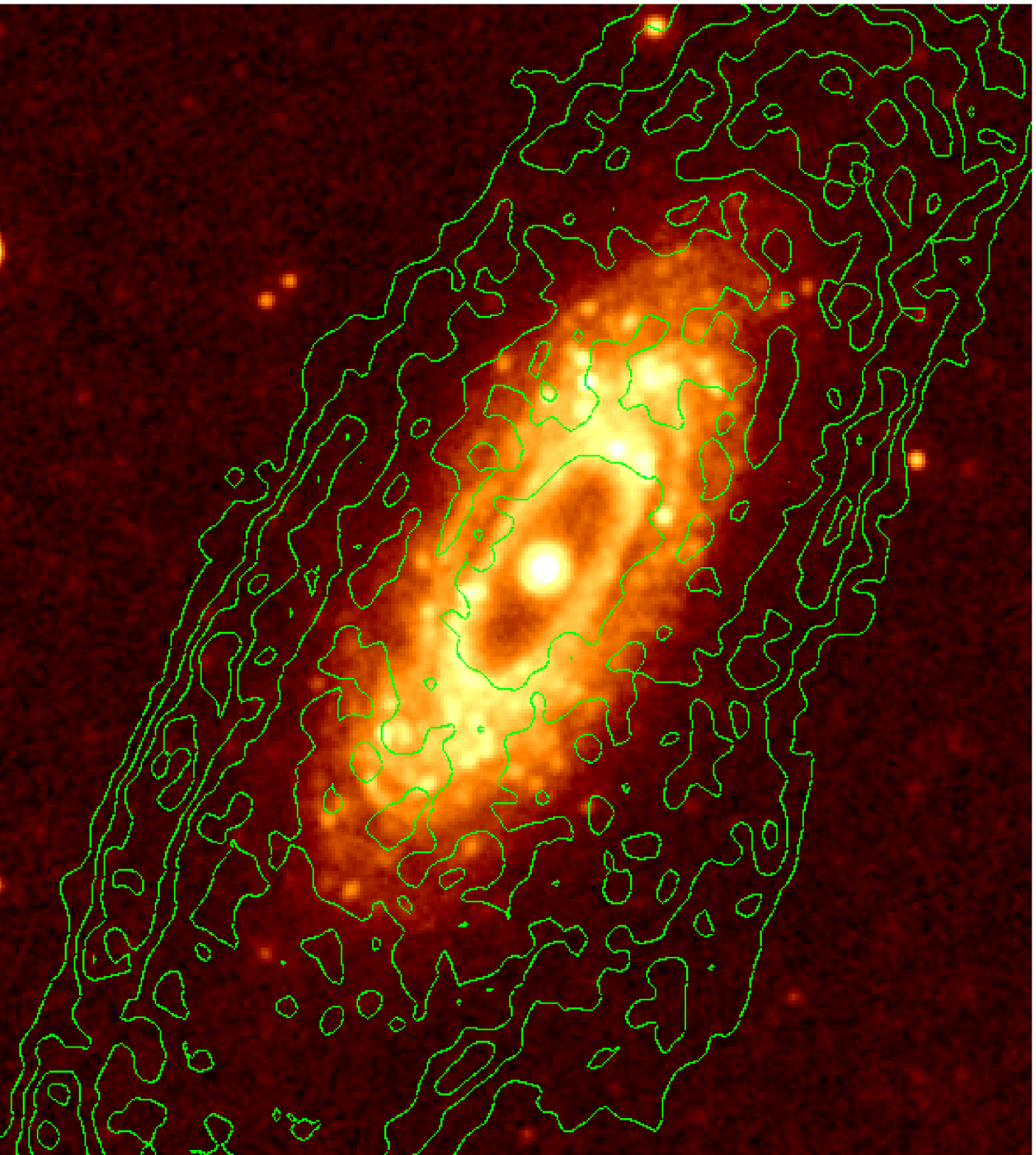}
      \caption{The $24\;\mu$m band image is plotted in color scale for the
       galaxies NGC~5194 (left) and NGC~2841 (right); the respective \hi\ 
       emission map is overlayed with green contours.}\label{fig:m51}
\end{figure*}

The present analysis is based on the 21~cm emission line maps, a tracer of the
neutral atomic gas for the 14 disk galaxies listed in Table~\ref{tab:objs},
which are taken from THINGS. These high-quality NRAO\footnote{The National
  Radio Astronomy Observatory is a facility of the National Science Foundation
  operated under cooperative agreement by Associated Universities, Inc.}\ Very
Large Array observations provide data cubes with an angular resolution of
$\simeq6''$ and spectral resolution of 2.6 or 5.2~\kms. Since the target
galaxies are nearby, at distances of 3--10~Mpc, the linear resolution of the
maps corresponds to 100--300~pc. The \hi\ data cubes of our target galaxies
are complemented with near-IR images, which are public data. In particular,
the majority of the THINGS galaxies (including all those in
Table~\ref{tab:objs}) have also been observed within the framework of the
SINGS and we make an extensive use of the $24\;\mu$m MIPS images (see
\S~\ref{sec:azcc}). Figure~\ref{fig:m51}\ illustrates our data for two of the
sample galaxies, NGC~5194 and NGC~2841. The $24\;\mu$m band image is shown in
color scale, and the contours show the \hi\ emission map. To obtain the
exponential scale length of the stellar disk (see \S~\ref{sec:analysis}), we
use $3.6\;\mu$m Infrared Array Camera (IRAC) images when available, otherwise
we use $H$ band images taken from the Two Micron All Sky Survey
\citep*[2MASS;][]{jarrett03}. To check the consistency of our results, we use
CO maps from the Berkeley-Illinois-Maryland Aaaociation Survey Of Nearby
Galaxies \citep*[BIMA-SONG][]{helfer03}\ for some of our target galaxies.

\begin{table*}
 \begin{center}
\caption{THINGS and SINGS Target Galaxies}  \label{tab:objs}
\begin{tabular}{lcccccccc}
\tableline
Obj. Name    & Alt. Name    &  $R_{25}$ ($^\prime$) & $R_{s}$ ($^\prime$) & Band  & $i$ ($^\circ$) & P.A. ($^\circ$) & $D$ (Mpc) & $v_{\rm max}$ (\kms)  \\
               & &  (1)                  &   (2)                  &  (3)   &(4) & (5) &  (6)  & (7)  \\
\tableline
NGC~2403 &       &       9.98            &   $1.30^\star  $       &  $H$   & 63 & 124 & 3.22  &  128 \\
NGC~2841 &       &       3.88            &   $0.92^\star  $       &  3.6   & 74 & 153 & 14.1  &  331 \\
NGC~3031 & M81   &       10.94           &   $3.63\pm0.2  $       &  3.6   & 59 & 330 & 3.63  &  256 \\
NGC~3184 &       &       3.62            &   $0.92\pm0.09 $       &  $H$   & 16 & 179 & 11.1  &  260 \\
NGC~3351 &       &       3.54            &   $0.86\pm0.03 $       &  3.6   & 41 & 192 & 9.33  &  210 \\
NGC~3521 &       &       4.8             &   $0.74\pm0.02 $       &  3.6   & 73 & 340 & 10.05 &  242 \\
NGC~3621 &       &       5.24            &   $0.80^\star  $       &  $H$   & 65 & 345 & 6.64  &  144 \\
NGC~3627 & M66   &       4.46            &   $0.95^\star  $       &  3.6   & 62 & 173 & 9.25  &  204 \\
NGC~5055 & M63   &       6.01            &   $1.16\pm0.05 $       &  $H$   & 59 & 102 & 7.82  &  209 \\
NGC~5194 & M51   &       3.88            &   $1.39\pm0.11 $       &  $H$   & 42 & 172 & 7.77  &  242 \\
NGC~628 &  M74   &       4.77            &   $1.10\pm0.09 $       &  $H$   & 7  & 20  & 7.3   &  220 \\
NGC~6946 &       &       5.35            &   $1.73\pm0.07 $       &  $H$   & 32.6 & 242 & 5.5 &  201 \\
NGC~7793 &       &       5.0             &   $1.16\pm0.05 $       &  $H$   & 50 & 290 & 3.82  &  109 \\
NGC~925  &       &       5.23            &   $1.43^\star  $       &  3.6   & 66 & 286 & 9.16  &  121 \\
\tableline
\end{tabular}
\tablecomments{Semi-major axis of the 25~mag~arcsec$^{-2}$ isophote in the $B$
  band obtained from the LEDA database (URL: http://leda.univ-lyon1.fr/); (2)
  exponential scale length derived in this paper as described in
  \S~\ref{sec:analysis}\ using either IRAF or {\tt galfit} (values tagged with
  $^\star$), where the error bars are $\delta R_s/R_s<1\;$\% ; (3) image band
  used (2MASS $H$ or IRAC $3.6\;\mu$m band) to derive $R_s$; (4) and (5)
  kinematic inclination and P.A., respectively; (6) adopted distance; (7)
  maximum amplitude of the rotation velocity corrected for inclination
  obtained from the rotation curve $v_c$ derived in \S~\ref{sec:diskmodel}.
  The values in Columns (4) to (6) are adopted from \citet{deblok07}.}
\end{center}
\end{table*}

\section{ANALYSIS}\label{sec:analysis}

All analysis in this paper started from fully reduced images and data cubes.
On this data we carry out two main steps. First, we derive the rotation curve
$v_c(r)$ of the \hi\ and the geometrical projection parameters of the galaxy
disk, and use these parameters to deproject the maps of the galaxies to
face-on orientation (see Table~\ref{tab:objs}). Second, we sample the face-on
maps in concentric annuli. For each annulus we cross-correlate the
corresponding pair of \hi\ and $24\;\mu$m fluxes, in order to derive the
angular offset between the \hi\ and the $24\;\mu$m patterns as a function of
radius.

For three of the galaxies listed in Table~\ref{tab:objs}\ (NGC~628, NGC~5194,
and NGC~3627), we also measure the angular offset between the CO and
$24\;\mu$m emission maps. If the ISM evolves sequentially from atomic into
molecular gas, and then subsequently initiates the formation of stars,
considering the kinematics expressed in Eq.~\ref{eq:offset}, we expect the CO
emission to lie in between the \hi\ and the $24\;\mu$m.

\subsection{Analysis of the \hi\ Kinematics}\label{sec:diskmodel}

For each object we apply the same general approach: we first perform adaptive
binning of the \hi\ data cube regions with low signal-to-noise (S/N) ratio
using the method described by \citet*{cappellari03}. From the resulting
spatially binned data cubes we fit the 21~cm emission lines with a single
Gaussian profile and use the parameterization to derive (1) the line-of-sight
velocity map $v(x,y)$, given by the line centroid, and (2) the flux maps
$\mu_0(x,y)\equiv a(x,y)/(\sqrt{2\pi}\,\sigma(x,y))$, where $a$ and $\sigma$
are the Gaussian peak amplitude and width, respectively. Since we do not need
to derive the rotation curve with high accuracy for the purpose of this paper,
we limit our model to a co-planar rotating disk with circular orbits described
by
\begin{equation}\label{eq:vcdisk}
v(x,y)=v_{\rm sys}+v_c(r)\,\sin i \cos\psi,
\end{equation}
where $v(x,y)$ is the observed velocity map along the line of sight
\citep*[see][]{begeman89}. For simplicity, we assume here that the orbits are
circular, though we address the issue of non-circular motions in
\S~\ref{sect:noncirc}. By $\chi^2$ minimization fitting\footnote{The fitting
  has been performed with the {\tt mpfit} IDL routine found at
  the URL:\\
  http://cow.physics.wisc.edu/$\sim${}craigm/idl/fitting.html\ } of the model
function in Eq.~\ref{eq:vcdisk}\ to the observed velocity map $v(x,y)$, we
obtain the systemic velocity $v_{\rm sys}=\rm const$, the inclination $i$ and
the position angle (P.A.) of the geometric projection of the disk on to the
sky.  Here, $\psi$ is the azimuthal angle on the plane of the inclined disk
(not the sky) and is a function of $i$ and P.A. The line where $\psi=0$
denotes the orientation of the line of nodes on the receding side of the disk.
The kinematic center $(x_0,y_0)$ is fixed a priori and is defined as the
central peak of either the IRAC $3.6\;\mu$m or the 2MASS $H$ band image. The
positions of the dynamical centers used here are consistent with those derived
in \citet*{trachternach07}.  We parameterize the deprojected rotation curve
$v_c$ with a four-parameter arctan-like function \citep*[e.g., ][]{rix97}
\begin{equation}\label{eq:vcrix}
v_c(r)=v_0\;(1+x)^\beta\; (1+x^{-\gamma})^{-1/\gamma},
\end{equation}
where $x=r/r_0$. Here, $r_0$ is the turn-over radius, $v_0$ is the scale
velocity, $\gamma$ determines the sharpness of the turnover and $\beta$ is the
asymptotic slope at larger radii.

The values for the projection parameters $i$, and P.A., the systemic velocity
$v_{\rm sys}$, and the asymptotic velocity that have been obtained applying
the approach described above, are consistent with the values reported in
Table~\ref{tab:objs}. From the maximum value of Eq.~\ref{eq:vcrix}\ we obtain
the maximum rotational velocity $v_{\rm max}$, which is listed for all the
sample galaxies in Table~\ref{tab:objs}.

\subsection{Azimuthal Cross-Correlation}\label{sec:azcc}

\begin{figure*}
\epsscale{0.8}
\plotone{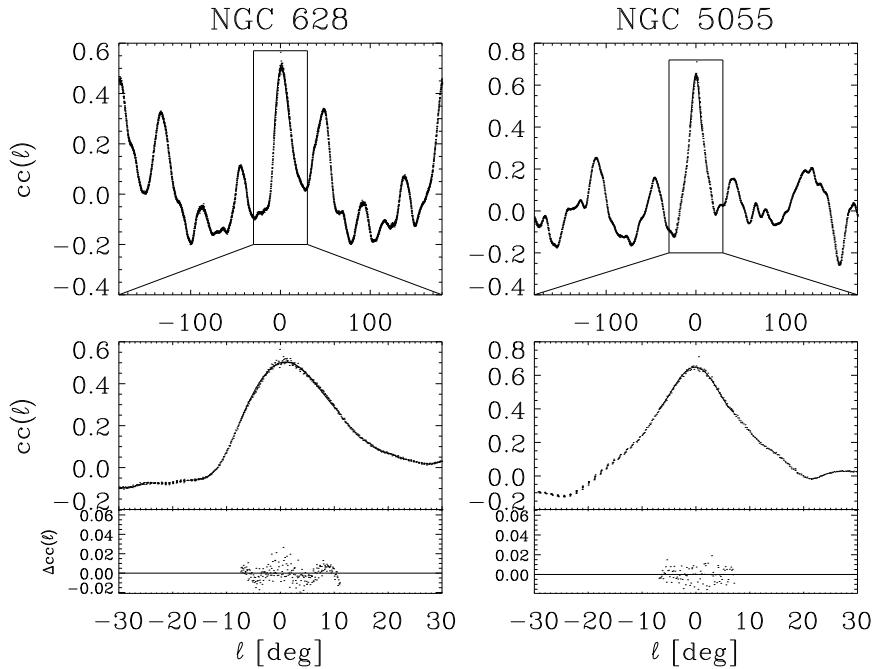}
     \caption{Representative examples for the determination of the azimuthal
       \hi--24~\mum\ offset: shown is the cross-correlation $cc(\ell)$ of the
       two functions $f_{\rm HI}(\phi_k,\hat{r})$ and $f_{24\,\mu{\rm
           m}}(\phi_{k-\ell},\hat{r})$, as a function of azimuth offset $\phi$
       at a fixed radius $\hat{r}$. The present example shows the $cc(\ell)$
       profile calculated for NGC~628 at $\hat{r}\simeq2'$ ($\simeq$4.2~kpc)
       and NGC~5055 at $\hat{r}\simeq2.6'$ ($\simeq$6.4~kpc), in the left and
       right columns, respectively.  Top panel: the $cc(\ell)$ profile in the
       entire range $[-180^\circ,180^\circ]$; bottom panel: a zoom of the
       range $[-30^\circ,30^\circ]$. We considered an adequate range greater
       than the width of the $cc(\ell)$ profile around $\ell_{\rm max}$ and
       interpolated $cc(\ell)$ locally ($\sim\pm10^\circ$ in the two example
       plots) with a fourth-degree polynomial
       $p_4(\ell)=\sum_{n=0}^4a_n\,\ell^n$, and calculated numerically the
       peak value $\ell_{\rm max}$. The bottom panel shows the fit residuals
       overplotted around the zero level.}\label{fig:reprex}
\end{figure*}

The central analysis step is to calculate by what angle $\Delta\phi$ the
patterns of \hi\ and $24\;\mu$m need to be rotated with respect to each other
in order to best match. We use the kinematically determined orientation
parameters, $i$ and P.A., to deproject both the \hi\ and $24\;\mu$m images to
face-on. To estimate the angular offset $\Delta\phi$ between the two flux
images at each radius, we divide the face-on images into concentric rings of
width $\sim$5$''$ and extract the flux within this annulus as a function of
azimuth. We then use a straightforward cross-correlation (CC) to search for
phase lags in $f_{\rm HI}(\phi|r)$ versus $f_{24\,\mu{\rm m}}(\phi|r)$. In
general, the best match between two discrete vectors $x$ and $y$ is realized
by minimizing as a function of the phase shift $\ell$ (also defined as {\it
  lag}) the quantity
\begin{equation}\label{eq:ccorrchi}
\chi^2_{x,y}(\ell)=\sum_k {\left[x_k-y_{k-\ell}\right]}^2,
\end{equation}
where the sum is calculated over all the $N$ elements of $x$ and $y$ with
$k=0,1,2,...,N-1$. Specifically here, for a given radius $r=\hat{r}$ we
consider for all discrete values of azimuth $\phi$:
\begin{equation}\label{eq:xydefs}
x_k=f_{\rm HI}(\phi_k|\hat{r})\quad\mbox{and}\quad 
y_{k-\ell}=f_{24\,\mu{\rm m}}(\phi_{k-\ell}|\hat{r}).
\end{equation}
Expanding the argument of the sum in Eq.~\ref{eq:ccorrchi}\ one obtains
that $\chi^2(\ell)$ is independent of the terms $\sum_k x_k^2$ and $\sum_k
y_{k-\ell}^2$, and $\chi^2$ is minimized by the maximization of
\begin{equation}\label{eq:ccorrdef}
cc_{x,y}(\ell)=\sum_k \left[x_k\,y_{k-\ell}\right],
\end{equation}
which is defined as the CC coefficient. Here we used the normalized CC
\begin{equation}\label{eq:ccorr}
cc_{x,y}(\ell)=\frac{\sum_k \left[(x_k-\bar{x})\,(y_{k-\ell}-\bar{y})\right]}
{\sqrt{\sum_k (x_k-\bar{x})^2\;\sum_k (y_k-\bar{y})^2}},
\end{equation}
where $\bar{x}$ and $\bar{y}$ are the mean values of $x$ and $y$ respectively.
Here the slow, direct definition has been used and not the fast Fourier
transform method.  The vectors are wrapped around to ensure the completeness
of the comparison.  With this definition the CC coefficient would have a
maximum value of unity for identical patterns, while for highly dissimilar
patterns it would be much less than 1.  We apply the definition in
Eq.~\ref{eq:ccorr}\ using the substitutions of Eq.~\ref{eq:xydefs}\ to compute
the azimuthal CC coefficient $cc(\ell)$ of the \hi\ and the $24\;\mu$m images.
The best match between the \hi\ and the $24\;\mu$m signals is realized at a
value $\ell_{\rm max}$ such that $cc(\ell_{\rm max})$ has its peak value.
Since the expected offsets are small (only a few degrees) we search the local
maximum around $\ell\simeq 0$. The method is illustrated in
Figure~\ref{fig:reprex}, which shows that $cc(\ell)$ has several peaks, as
expected due to the self-similarity of the spiral pattern.

We consider a range that encompasses the maximum of the $cc(\ell)$ profile,
i.e. the central $\sim$100--150 data points around $\ell_{\rm max}$.  This
number, depending on the angular size of the ring, is dictated by the
azimuthal spread of the spiral arms and the number of substructures (e.g.,
dense gas clouds, star clusters, etc.) per unit area.  This corresponds, for
example, to a width in $\ell$ of a few tens of degrees at small radii ($\sim
1'$), depending on the distance of the object, and a range in width of $\ell$
decreasing linearly with the radius. We interpolate $cc(\ell)$ around
$\ell_{\rm max}$ with a fourth-degree polynomial using the following
approximation: $cc(\ell)\simeq p_4(\ell)=\sum_{n=0}^4a_n\,\ell^n$ and
calculate numerically (using the Python\footnote{http://www.python.org\ }
package {\tt scipy.optimize}) the peak value at $\ell_{\rm max}$,
$p_4(\ell_{\rm max})$. By repeating the procedure for all radii, the angular
offset \hi$\mapsto$24$\;\mu$m results in $\Delta\phi(r)=-\ell_{\rm max}(r)$.
The direction or equivalently the sign of the lag $\ell_{\rm max}$ between two
generic vectors $x$ and $y$ depends on the order of $x$ and $y$ in the
definition of the CC coefficient.  Note that $cc_{x,y}(\ell)$ in
Eq.~\ref{eq:ccorrdef}\ is not commutative for interchange of $x$ and $y$,
being $cc_{y,x}(\ell)=cc_{x,y}(-\ell)$. For $\ell_{\rm max}=0$ the two
patterns best match at zero azimuthal phase shift. The error bars for
$\delta\ell_{\rm max}(r)$ have been evaluated through a Monte Carlo approach,
adding normally distributed noise and assuming the expectation values of
$\ell_{\rm max}$ and $\delta\ell_{\rm max}$ as the mean value and the standard
deviation, respectively, after repeating the determination $N=100$ times.

\begin{figure*}
\epsscale{1.1}
\plotone{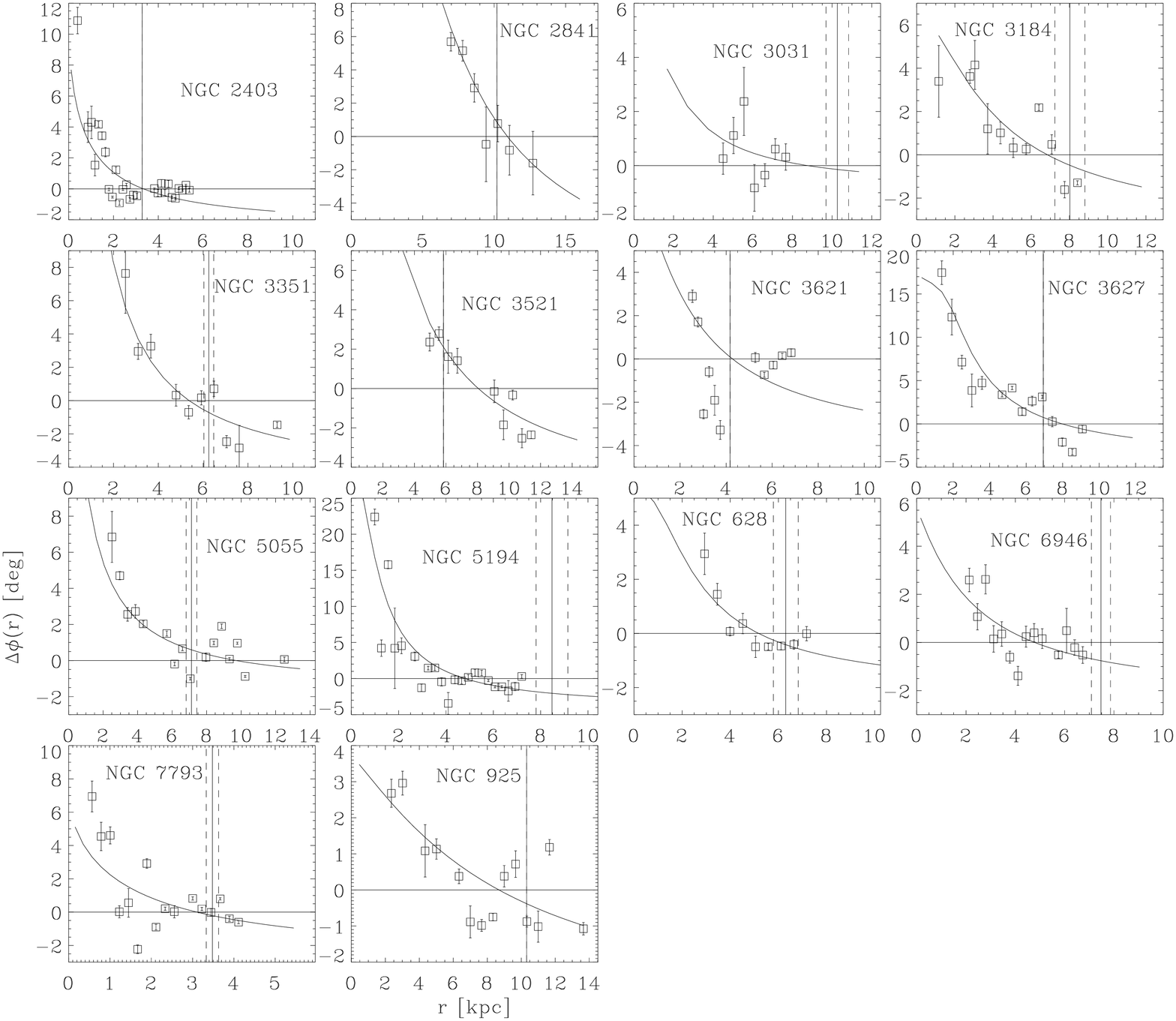}
    \caption{Radial profiles for the angular offset \hi $\mapsto$24$\;\mu$m
      for the entire sample, obtained by sampling face-on \hi\ and 24$\;\mu$m
      maps concentric rings and cross-correlating the azimuthal profiles
      for each radius. The solid line is the best-fit model to the observed
      data points, denoted by squared symbols, which has been obtained by
      $\chi^2$ minimization of Eq.~\ref{eq:offsetdeg}; the solid curve
      intersects the horizontal axis at corotation (defined as
      $\Delta\phi=0$). The solid and dashed vertical lines indicate the
      $2.7\:R_s\simeq R_{\rm cor}$ value and error bars, derived by
      \citet*{kranz03}.}\label{fig:dlr}
\end{figure*}

Our analysis is limited to the radial range between low S/N regions at the
galaxy centers and their outer edges.  In the \hi\ emission maps the S/N is
low near the galaxy center, where the \hi\ is converted to molecular H$_2$,
whereas for the $24\;\mu$m band the emission map has low S/N near $R_{25}$
(and in most cases already at $\sim0.8\:R_{25}$).  Regions with S/N~$<3$ in
either the \hi\ or $24\;\mu$m images have been clipped. We also ignore those
points $\ell_{\rm max}$ with a coefficient $cc(\ell_{\rm max})$ lower than a
threshold $cc\simeq0.2$.  We further neglect any azimuthal ring containing
less than a few hundred points, which occurs near the image center and near
$R_{25}$. The resulting values $\Delta\phi(r)$ are shown in
Figure~\ref{fig:dlr}.

\subsection{Disk Exponential Scale Length}
We also determine the disk exponential scale length $R_s$ for our sample
using the {\tt galfit}\footnote{Found at URL:\\
  http://zwicky.as.arizona.edu/$\sim$cyp/work/galfit/galfit.html\ } algorithm
\citep*{peng02}. In particular, we fit an exponential disk profile and a de
Vaucouleurs profile to either the IRAC $3.6\;\mu$m or to the 2MASS $H$ band
image. As {\tt galfit} underestimates the error on $R_s$ (as recognized by the
author of the algorithm), typically $\delta R_s/R_s<1$\%, we therefore also
use the IRAF task {\tt ellipse} \citep*{jedrzejewski87}\ to derive the radial
surface brightness profile and fit $R_s$. After testing the procedure on a few
objects, we note only small differences (of the order of the error bars in
Table~\ref{tab:objs}) when deriving $R_s$ from the $H$ band and the
$3.6\;\mu$m band.

\section{RESULTS}\label{sec:results}

\subsection{Angular Offset}\label{sec:results_ang}

With the angular offset $\Delta \phi(r)\equiv \langle\phi_{24\,\mu{\rm
    m}}-\phi_{\rm HI}\rangle(r)$, where $\phi$ increases in the direction of
rotation, and the rotation curve $v_c(r)$ for each radial bin, we can rewrite
Eq.~\ref{eq:offset}\ as
\begin{equation}\label{eq:offsetdeg}
\Delta \phi(r) =\left(\frac{v_c(r)}{r}-\Omega_p\right)
\;\times\;t_{{\rm HI}\mapsto 24\,\mu{\rm m}},
\end{equation}
where $\Omega(r)\equiv v_c/r$. Since $\Omega(r)>\Omega_p$ inside the
corotation radius $R_{\rm cor}$, and $\Omega(r)<\Omega_p$ outside corotation,
we expect $\Delta\phi(r)> 0$ for $r<R_{\rm cor}$ and $\Delta\phi(r)< 0$ for
$r>R_{\rm cor}$. At corotation, where $\Delta\phi(R_{\rm cor})= 0$, the two
components \hi\ and $24\;\mu$m should have no systematic offset.  We assume
\tHI\ and $\Omega_{\rm p}$ to be constant for any given galaxy, and that all
the spirals are trailing, since the only spiral galaxies known to have a
leading pattern are NGC~3786, NGC~5426 and NGC~4622
\citep*{thomasson89,byrd02}.  By $\chi^2$ fitting the model prediction of
Eq.~\ref{eq:offsetdeg}\ to the measured angular offsets $\Delta\phi(r)$ in all
radial bins of a galaxy, we derive best-fit values for \tHI\ and $\Omega_{\rm
  p}$.

The $\Delta\phi(r)$ data and the resulting best-fits are shown in
Figure~\ref{fig:dlr}, with the resulting best-fit values listed in
Table~\ref{tab:omet}. In Figure~\ref{fig:dlr}\ we plot for all objects the
radial profile of the angular offsets $\Delta\phi(r)$. The solid line
represents the best fit model proscribed by Eq.~\ref{eq:offsetdeg}. The square
symbols in the plot represent the fitted data points from \S~\ref{sec:azcc}.
Looking at the ensemble results in Figure~\ref{fig:dlr}, two points are
noteworthy: (1) the geometric offsets are small, typically a few degrees and
did need high-resolution maps to become detectable, (2) the general radial
dependence follows overall the simple prescription of Eq.~\ref{eq:offsetdeg}\ 
quite well.

\begin{table*}
\begin{center}
\caption{Characteristic Timescales \tHI\ and Pattern Speed $\Omega_p$
  Resulting from a $\chi^2$ fit of the Observed Angular Offset via
  Equation~\ref{eq:offset}  } \label{tab:omet}
\begin{tabular}{lccccc}
\tableline
 Obj. Name& Alt. Name & \tHI  & $\Omega_p$ &
 $R_{\rm cor}/R_s$ & $\Omega_p$ \\
 & &  [Myr] &  [km s$^{-1}$ kpc$^{-1}$] & & [km s$^{-1}$ kpc$^{-1}$] \\
\tableline
NGC~2403  &                & $1.4\pm0.5$  & $30\pm4$   & $2.8\pm0.3$ & \\
NGC~2841  &                & $4.4\pm0.5$  & $42\pm2$  & $2.8\pm0.1$ & \\
NGC~3031  & M81           & $0.5\pm0.3^\dag$  & $27\pm13^\dag$  & $2.3\pm1.4$  & $24^{\rm a}$ \\
NGC~3184  &                & $1.8\pm0.4$  & $38\pm5$   & $2.3\pm0.5$ & \\
NGC~3351  & M95           & $2.2\pm0.3$  & $38\pm3$  & $2.3\pm0.4$  & \\
NGC~3521  &                & $2.9\pm0.4$  & $32\pm2$  & $3.6\pm0.2$ & \\
NGC~3621   &               & $2.3\pm1.3^\dag$  & $31\pm11^\dag$  & $2.8\pm1.0$ & \\
NGC~3627  & M66           & $3.1\pm0.4$  & $25\pm4$   & $3.0\pm0.5$  & \\
NGC~5055  & M63           & $1.3\pm0.3$  & $20\pm5$   & $3.8\pm1.5$  & $30-40^{\rm b}$ \\
NGC~5194  & M51           & $3.4\pm0.8$  & $21\pm4$   & $1.5\pm0.6$  & $38\pm7^{\rm c}$, $40\pm8^{\rm d}$ \\
NGC~628  & M74            & $1.5\pm0.5$  & $26\pm3$   & $2.2\pm0.4$  &  $32\pm2^{\rm e}$ \\
NGC~6946  &                & $1.3\pm0.3^\dag$  & $36\pm4^\dag$   & $1.7\pm0.4$ & $39\pm9^{\rm c}$, $42\pm6^{\rm d}$ \\
NGC~7793  &                & $1.2\pm0.5^\dag$  & $40\pm10^\dag$  & $2.5\pm0.9$ &  \\
NGC~925  &                 & $5.7\pm1.6$  & $11\pm1.0$ & $2.1\pm0.2$ & $7.7^{\rm f}$ \\
\tableline
\end{tabular}
\tablecomments{The error bars are evaluated via a Monte Carlo method. The
  timescales \tHI\ listed in this table are summarized in the histogram in
  Figure~\ref{fig:hist}. The corotation radius to exponential scale radius
  $R_{\rm cor}/R_s$ ratios are summarized in Figure~\ref{fig:rco}. We list for
  comparison in the last column other measurements of the pattern speed: (a)
  \citet*{westpfahl91}; (b) \citet*{thornley1997}; (c) \citet*{zimmer04}; (d)
  \citet*{hernandez04}; (e) \citet{Sakhibov2004}; (f) \citet*{elmegreen98}.
  Addressing to the discussion on pattern speeds in \S~\ref{sec:pspeeds}, we
  tag as bad fit those galaxies (indicated with the $\dag$ symbol) where the
  fit is not reliable.}
\end{center}
\end{table*}

\subsection{\tHI\ and $R_{\rm cor}$}\label{sec:corotation}

Because $\Delta\phi(r)$ is consistent with (and follows) the predictions of
the simple geometry and kinematics in Eq.~\ref{eq:offsetdeg}, the procedure
adopted here turns out to be an effective method to derive the following: (1)
the time lag \tHI, which should bracket the timescale needed to compress the
molecular gas, trigger star formation, and heat the dust; it therefore
represents also an estimate for the lifetime of star-forming molecular clouds;
(2) the kinematic pattern speed $\Omega_p$ of the galaxy spiral pattern and,
equivalently, the corotation radius $R_{\rm cor}$.
  
We now look at the ensemble properties of the resulting values for \tHI\ and
$R_{\rm cor}$. The scatter of the individually fitted $\Delta\phi$ points is
significantly larger than their error bars (as shown in Fig.~\ref{fig:dlr}),
which may be due to the galactic dynamics being more complex than our simple
assumptions.  For example, the pattern speed may not be constant over the
entire disk or there may be multiple corotation radii and pattern speeds, as,
for instance, found by numerical simulations \citep*{sellwood88}\ and observed
in external galaxies \citep*{hernandez05}.

\begin{figure}
\epsscale{1.0}
\plotone{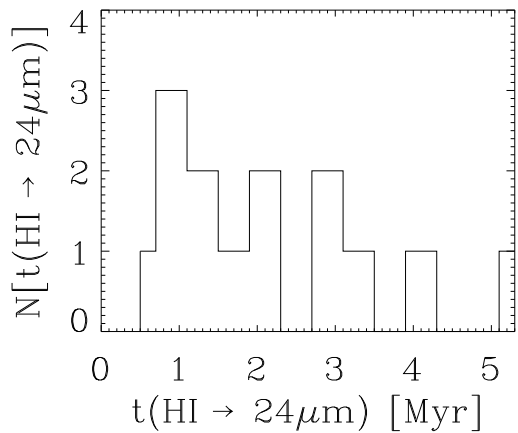}
     \caption{Histogram of the timescales \tHI\ derived for the 14
       sample galaxies listed in Table~\ref{tab:objs}\ from the fits in
       Figure~\ref{fig:dlr}\ (also listed in Table~\ref{tab:omet}). The
       timescales range between 1 and 4~Myr for almost all
       galaxies.}\label{fig:hist}
\end{figure}

Even though a considerable intrinsic scatter characterizes $\Delta\phi(r)$ for
most of our sample galaxies, and the error bars of \tHI\ listed in
Table~\ref{tab:omet}\ are typically $>15\,$\%, the histogram of the
characteristic timescales \tHI\ in Figure~\ref{fig:hist}\ shows overall a
relatively small spread for a sample of 14 galaxies of different Hubble types:
the timescales \tHI\ occupy a range between 1 and 4 Myr for almost all the
objects.

\begin{figure}
\plotone{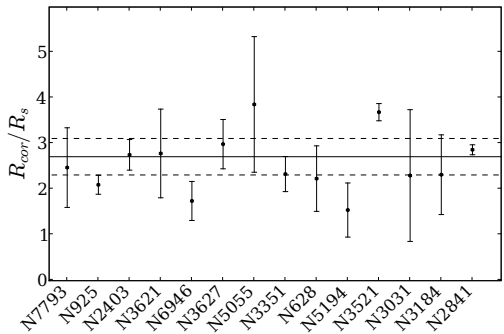}
     \caption{The best fit for the spiral arm corotation radius (in units of
       exponential scale radii) obtained by inverting the best-fit model of
       Eq.~\ref{eq:offset}\ at $\Delta\phi(r)=0$, and evaluating $R_s$ via a
       bulge-disk decomposition on either $H$ band or $3.6\;\mu$m band images.
       The solid and dashed horizontal lines represent the $R_{\rm
         cor}/R_s=2.7\pm0.4$ value found by \citet*{kranz03}. The galaxies in
       the plot are sorted by asymptotic rotation velocity (see
       Table~\ref{tab:objs}), giving no indication of a correlation
       between dynamical mass and the $R_{\rm cor}/R_s$ ratio.}\label{fig:rco}
\end{figure}

The solid curve in Figure~\ref{fig:dlr}, representing the prescription of
Eq.~\ref{eq:offsetdeg}, intersects the horizontal axis at the corotation
radius $R_{\rm cor}$, which can be formally derived by inverting
Eq.~\ref{eq:offsetdeg}\ at $\Delta\phi=0$. We report in Table~\ref{tab:omet}\ 
the ratio between $R_{\rm cor}$ and the exponential scale length $R_s$ for
each object and show this result in Figure~\ref{fig:rco}. Comparisons of our
pattern speed $\Omega_p$ measurements with other methodologies
\citep*[e.g.][]{tremaine84}\ are listed in Table~\ref{tab:omet}. The
differences with our results may arise since the Tremaine--Weinberg method
assumes the continuity condition of the tracer, which may break down for the
gas as it is easily shocked, it changes state, and it is converted into stars
\citep{hernandez05,rand04}, or it can be obscured by dust \citep{gerssen07}.
Comparison of the observed non-axisymmetric motions with hydrodynamical models
based on the actual stellar mass distribution \citep*{kranz03}\ have found a
characteristic value of $R_{\rm cor}/R_s \simeq 2.7\pm0.4$ for a sample of
spirals. We plot this range with dashed vertical lines in
Figure~\ref{fig:dlr}, whereas the solid curve in each panel denotes the actual
fit value of $R_{\rm cor}/R_s$ for each object.  Figure~\ref{fig:rco}\ 
illustrates the interesting fact that the corotation values found in the
present paper, $R_{\rm cor}/R_s\simeq2.7\pm0.2$, agree well with the
completely independent estimates that \citet*{kranz03} derived by a different
approach and for a different sample. Therefore, a generic value for the
pattern speed of the dominant spiral feature of $R_{\rm cor}/R_s \simeq
2.7\pm0.4$ seems robust.

\subsection{Comparison With CO Data}

\begin{figure}
\plotone{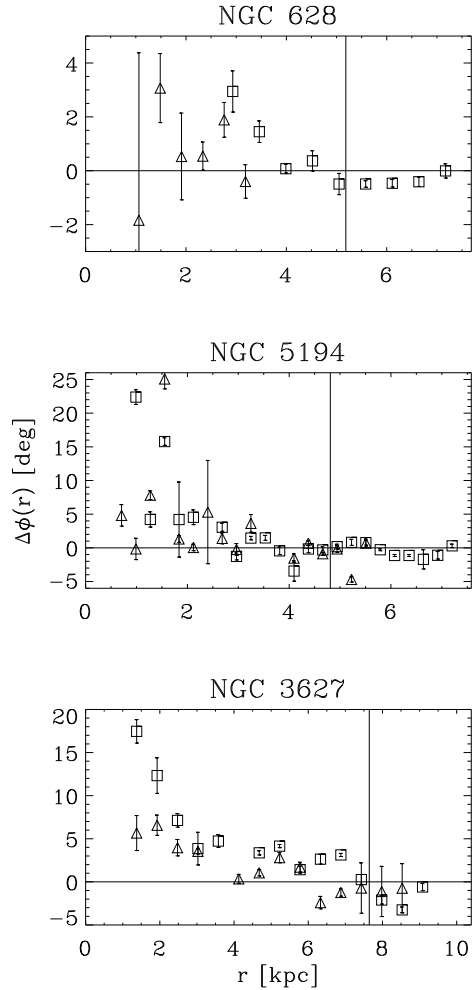}
     \caption{Comparison of the angular offsets obtained for
       \hi$\mapsto$24$\;\mu$m (squares) and CO$\mapsto$24$\;\mu$m (triangles)
       for the galaxies NGC~628, NGC~5194, and NGC~3627. CO maps are taken
       from the BIMA-SONG survey.  The solid vertical line in each panel
       indicates the position of corotation as obtained by $\chi^2$ fitting of
       Eq.~\ref{eq:offset}\ for the \hi. These results are qualitatively
       consistent with a temporal star-formation sequence
       \hi$\mapsto$CO$\mapsto$24$\;\mu$m.}\label{fig:HICO}
\end{figure}

If the basic picture outlined in the introduction is correct, then the
molecular gas traced by the CO, as an intermediate step in the star-formation
sequence, should lie in between and have a smaller offset from the \hi\ than
the $24\;\mu$m does, but in the same direction. To check this qualitatively,
we retrieve the BIMA-SONG CO maps \citep*{helfer03}\ for the galaxies NGC~628,
NGC~5194, and NGC~3627. For comparison, we derive the angular offset between
the CO emission and the $24\;\mu$m, applying the same method described above
for the \hi. The results are plotted in Figure~\ref{fig:HICO}.  The scarcity
of data points (e.g., NGC~628) and their scatter, which is typically larger
than the error bars, make estimates of $t_{{\rm CO}\mapsto 24\,\mu{\rm m}}$
and $R_{\rm cor}$ rather uncertain.  Therefore, we simply focus on
$\Delta\phi_{\rm CO-24}(r)$ versus $\Delta\phi_{\rm {\sc Hi}-24}(r)$, which is
shown in Figure~\ref{fig:HICO}.  This figure shows that the values of
$\Delta\phi_{\rm CO-24}$ all lie closer to zero than the values of
$\Delta\phi_{\rm {\sc Hi}-24}$.

Hence this check shows that the peak location of the molecular gas is
consistent with the evolutionary sequence where the \hi\ represents an earlier
phase than the CO. In this picture the \hi\ has a larger spatial separation
with respect to the hot dust emission, except at corotation, where the three
components are expected to coincide. However, it is clear that higher
sensitivity CO maps are needed to improve this kind of analysis.

\subsection{Analysis of Non-Circular Motions}\label{sect:noncirc}

\begin{figure}
\plotone{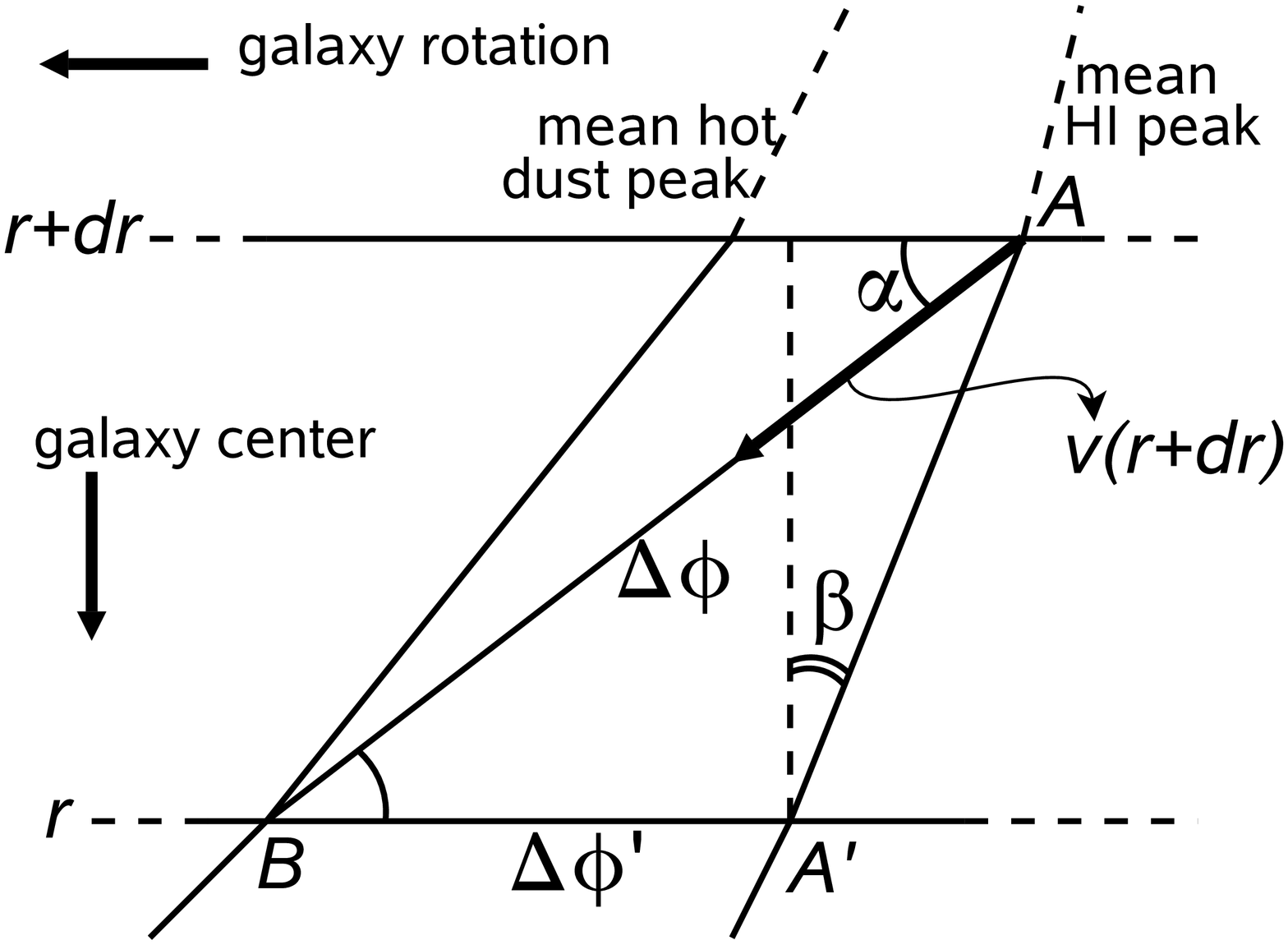}
\caption{Accounting for non-circular motions in the frame corotating with the
  spiral pattern, the two horizontal lines denoted with $r$ and $r+\der r$
  represent two galactocentric distances. Gas (and the resulting young stars)
  move along the line $\overline{AB}\propto\Delta\phi$, proceeding from the
  mean location of dense \hi\ to the mean 24$\;\mu$m peak, both denoted with
  solid lines, where $\beta$ is the \hi\ pitch angle. The  gas velocity
  vector is at an angle $\alpha$ with respect to the circle at radius $r+\der
  r$.   The line $\overline{A'B}$ denotes the (biased) measure of
  $\Delta\phi^\prime$ at constant radius $r$, which we actually estimate in
  our data analysis, and which is shorter than the actual shift value
  $\overline{AB}\propto\Delta\phi$ by a factor of
  $\cos\beta/\cos(\beta+\alpha)$, as explained in the
  text.}\label{fig:noncartoon}
\end{figure}

So far we have carried out an analysis that is based on the assumption of
circular motions.  We quantify here non-circular motions and determine to what
extent their presence affects the estimate of the timescales \tHI, which scale
with $\Delta\phi$. In the classic picture (e.g., R69), the radial velocity of
the gas is reversed around the spiral shock, so that the material is at nearly
the same galactocentric radius before and after the shock. Gas in galaxies
with dynamically important spiral arms does not move on circular orbits,
though. Shocks and streaming motions transport gas inwards, and gas orbits
undergo strong variations of direction. If a continuity equation for a
particular gas phase applies, it implies that in the rest frame of the spiral
arm, the change of relative velocity perpendicular to the arm $v_\perp$ is
proportional to the arm to pre-arm mass flux ratio \citep[for a recent
illustration in M51 see][ hereafter S07]{shetty07}. For example, for an orbit
passing through an arm with mass density contrast of 10, $v_\perp$ would drop
by the same factor, producing a net inward deflection of the orbit. We do note
that the gas continuity equation may not actually be valid, since stars may
form, or the gas may become molecular or ionized. Non-circular motions could
modify the simplified scheme of Fig.~\ref{fig:cartoon}. If the orbit is inward
bound near the arm, the path of the material between the \hi\ and the
24$\;\mu$m arm components is larger than that previously assumed for circular
orbits. We reconsider the scheme of Fig.~\ref{fig:cartoon}\ for a non-circular
orbit in the frame corotating with the spiral pattern as illustrated in
Fig.~\ref{fig:noncartoon}. Here, the material moves not along a line at
constant radius $r$, but along a line proceeding from a larger radius $r+\der
r$, specifically from the point $A$ (see Fig.~\ref{fig:noncartoon}) on the
\hi\ arm toward the point $B$ on the 24$\;\mu$m arm, where the two parallel
horizontal lines denote the galactocentric distances $r+\der r$ and $r$. The
material departs from $A$ with an angle $\alpha$ inwards (if the material were
to proceed instead from an inner radius, then $\alpha$ is directed outwards).
If $\alpha$ is large, the measurement of the spatial shift
$\overline{A'B}\simeq r\,\Delta\phi^\prime$ in the simple scheme of circular
(deprojected) rings no longer represents the actual value of the spatial
offset $\overline{AB}\propto\Delta\phi$, but rather is only a lower limit.
Consequently, the measurement of the timescale \tHI\ would be underestimated,
since \tHI~$\propto\Delta\phi$ from Eq.~\ref{eq:offsetdeg}.  From the geometry
of the triangle $ABA'$ in Fig.~\ref{fig:noncartoon}, and since
$\overline{A'B}\propto\Delta\phi^\prime$ and $\overline{AB}\propto\Delta\phi$,
it can be straightforwardly shown that
\begin{equation}\label{eq:offset_corr}
  \Delta\phi(r)=\Delta\phi^\prime(r)\:\frac{\cos\beta}{\cos(\beta+\alpha)},
\end{equation}
where 
\begin{equation}\label{eq:tanalpha} 
\tan\alpha=\left. \frac{-v_R}{v_\phi^\prime} \right|_{r+\der r}, 
\end{equation} 
$v_R(r)$ and $v_\phi^\prime(r)=v_\phi-\Omega_p\,r$ are the radial and
tangential velocity components of the gas in the arm frame, respectively, and
$\beta$ is the \hi\ arm pitch angle defined by
\begin{equation}\label{eq:pitchangle} 
\tan\beta= \left. \frac{\der\phi_{\rm arm}}{\der r}
\right|_{r}, 
\end{equation} 
with $\beta\rightarrow90^\circ$ for a tightly wound spiral. Note that the
radial and the azimuthal dependences of $v_R$ and $v_\phi^\prime$ are anchored
to each other at the spiral arms through Eq.~\ref{eq:pitchangle}.  It follows
that if the radial and tangential components of the gas velocity, and the
pitch angle of the \hi\ spiral arms are known, the actual value of
$\Delta\phi$ can be calculated by applying the correction factor
\begin{equation}\label{eq:correction}
k(r)\equiv\frac{\cos\beta}{\cos(\beta+\alpha)}
\end{equation}
to the directly measured quantity $\Delta\phi^\prime$. Note from
Fig.~\ref{fig:noncartoon}\ that if $\alpha\rightarrow90-\beta$, then
$k(r)\rightarrow\infty$, but in this case also
$\Delta\phi^\prime\rightarrow0$, resulting in a finite value for $\Delta\phi$
through Eq.~\ref{eq:offset_corr}, and for the timescale, since
\tHI~$\propto\Delta\phi$.

\subsubsection{Methodology}

In these circumstances, if spiral arms in galaxies are dynamically affecting
the gas kinematics, then it is sensible to test the effects of non-circular
motions especially in strong arm spiral galaxies from our sample. We select,
NGC~5194 (M51), NGC~628, and NGC~6946.  Therefore, we examine the extreme
case, calculating along a narrow region along spiral arms how much the
streaming motions affect our angular offset measurements, thus the derivation
of the timescales \tHI.  Subsequently, we generalize the procedure accordingly
with our specific methodology by averaging these effects over the product of
\hi~$\times$~24~\mum\ fluxes, which is the weighting function of the
cross-correlation.

\begin{figure}
\epsscale{1.0}
\plotone{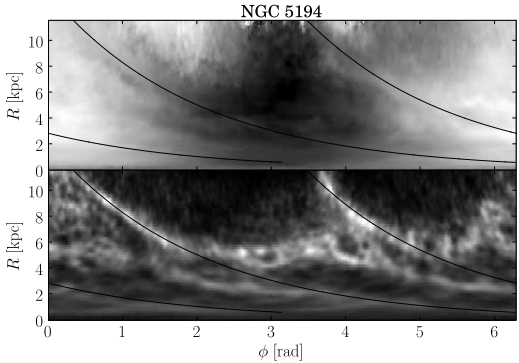}
\caption{The figure illustrates logarithmic spirals of
  the form $\psi= \phi_{\rm arm}-\phi_0=\ln (R_{\rm arm}/R_0)$ where the
  observed velocity $v_{\rm obs}$ is extracted in order to be fitted to
  Eq.~\ref{eq:generalv}\ and obtain the $v_R$ and $v_\phi$ velocity components
  (see \S~\ref{sect:noncirc}).  The top and bottom panels represent the
  projection 
  into a polar coordinates system ($R,\phi$) of the observed \hi\ 
  line-of-sight velocity field and the \hi\ column density map, respectively,
  for the galaxy NGC~5194. The solid lines represent logarithmic spiral arms
  with phases $\psi=\phi_{\rm arm}$ and $\psi=\phi_{\rm arm}\pm180$\dg. In the
  top panel, the gray scale image of the velocity field indicates velocity
  values from $-90$~\kms\ (dark) to $+90$~\kms\ (light). The velocity map in
  figure is not deprojected for inclination effects, which are instead taken
  into account in Eq.~\ref{eq:generalv}.  The coordinate $\phi=0$ represents
  the kinematic position angle of the galaxy.}\label{fig:polarv}
\end{figure}

The radial and tangential velocity components are difficult to separate
unambiguously. The radial component is accurately measured along the minor
axis, where however information on the azimuthal component is lost. The
opposite occurs along the major axis.  Therefore, we follow the prescriptions
for non-circular streaming motions analysis from \S~3 of S07, to separate the
velocity components $v_R$ and $v_\phi$ from the observed \hi\ velocity field
$v_{\rm obs}$, adopting
\begin{equation}
  \label{eq:generalv}
  v_{\rm obs}(R,\phi)=v_{\rm sys} + (v_R(R,\phi)\,\sin\phi+v_\phi(R,\phi)\,\cos\phi) \:\sin i
\end{equation}
as a generalization of Eq.~\ref{eq:vcdisk}. Knowing $v_R$ and $v_\phi$, then
we obtain the geometry of the orbits using Equations~\ref{eq:tanalpha}\ and
\ref{eq:pitchangle}. Specifically, the \hi\ column density and velocity map
are deprojected and resampled into a polar coordinate system ($R,\phi$) for
simplicity, so that the azimuthal phase of the spiral arms can be described by
a logarithmic spiral $\psi= \phi_{\rm arm}-\phi_0=\ln (R_{\rm arm}/R_0)$, for
a given fiducial radius $R_0$.  We extract the observed velocity $v_{\rm obs}$
along logarithmic lines as illustrated in Fig.~\ref{fig:polarv}, which best
represent the arm phase. Though these logarithmic lines are drawn by eye, on
top of the spiral arms, we obtain reasonable results, e.g., for the case of
NGC~5194 we find a logarithmic slope of 26\dg, similar to the 21\dg\ found by
S07.  Assuming $v_R$ and $v_\phi$ to be constant along equal arm phases, we
fit Eq.~\ref{eq:generalv}\ to the observed velocities extracted at each arm
phase $\psi\in[\phi_{\rm arm},\phi_{\rm arm}+2\pi]$. The fitting gives $v_R$
and $v_\phi$ as a function of $\psi$ and therefore as a function of radius
through Eq.~\ref{eq:pitchangle}\ (see also S07).  Provided the angles $\alpha$
and $\beta$ from Eq.~\ref{eq:tanalpha}\ and Eq.~\ref{eq:pitchangle}, we
straightforwardly calculate the actual value of the angular offset
$\Delta\phi$ given $\Delta\phi^\prime$ from circular orbits assumption using
Eq.~\ref{eq:offset_corr}\ and Eq.~\ref{eq:correction}.

\begin{figure}
\plotone{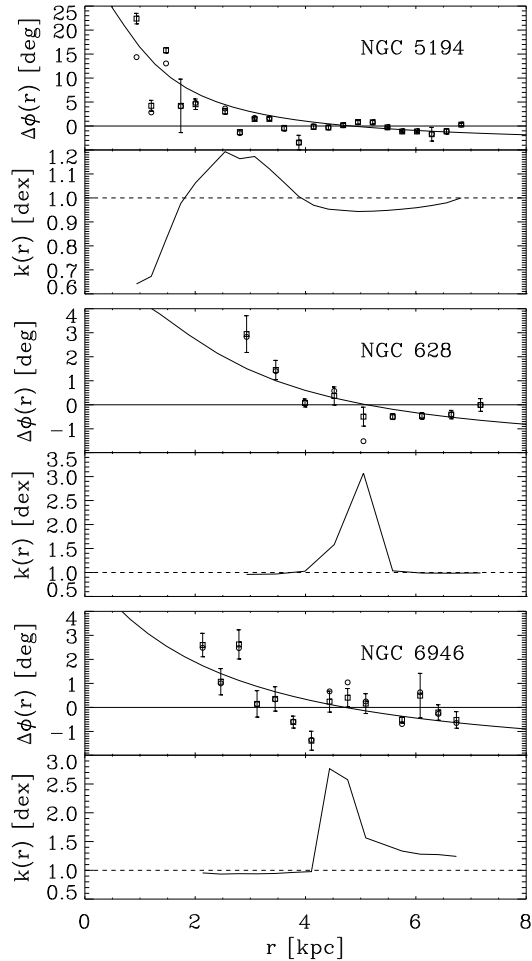}
 \caption{Correction factor $k(r)$ from Eq.~\ref{eq:correction}, as applied to
   the angular offset measurement for the galaxies NGC~5194, NGC~628, and
   NGC~6946.  Bottom panels: the solid curve represents the correction factor
   $k(r)=\cos\beta/\cos(\beta+\alpha)$ as a function of radius. Top panels:
   the squares denote the angular offset measurements $\Delta\phi^\prime$ of
   Fig.~\ref{fig:dlr} calculated assuming circular orbits, the circles denote
   $\Delta\phi$ after correction, and the solid curve represents the model fit
   of Eq.~\ref{eq:offsetdeg}\ to the corrected offset
   values.}\label{fig:correction}\end{figure}

Consistently with \citet{gomez02}\ and S07, we find in the three considered
galaxies that the locations of the spiral arms coincide with a net drop-off of
the tangential velocity and negative radial velocity, possibly indicating that
near the arms the orbits bend inwards.  We then calculate, using
Eq.~\ref{eq:correction}, the correction $k(r)$ at the position of the arms,
indicated in Fig.~\ref{fig:polarv}, where we expect the largest variations for
$v_R$ and $v_\phi$.  Even so, we find for the three considered galaxies that
$k(r)$ results near unity for all radii, except where $v_\phi^\prime$
approaches zero, which does not necessarily coincide with corotation, but
rather where the division in Eq.~\ref{eq:tanalpha}\ diverges and the error
bars are large.  In particular, for NGC~5194, where $|\alpha|<20^\circ$ for
all $r$, the correction $k(r)$ ranges between 0.7 and 1.2, and for both
NGC~628 and NGC~6946 $k(r)$ ranges between 0.9 and 1.5, as indicated in
Fig.~\ref{fig:correction}.  After calculating the corrected offsets
$\Delta\phi$ from Eq.~\ref{eq:offset_corr}, and fitting
Eq.~\ref{eq:offsetdeg}\ to the values $\Delta\phi$, we find that the timescale
\tHI\ and the pattern speed $\Omega_p$ do not change significantly---the
differences are below the error bars for the three galaxies. Note that the
data points where $v_\phi^\prime\simeq0$, that is where Eq.~\ref{eq:tanalpha}\ 
diverges, are not excluded from the fit. The results of these fits are listed
in Table~\ref{tab:corromet}\ and plotted in Fig.~\ref{fig:correction}. We also
find that the radial displacements,
\begin{equation} \der r\simeq
  r\,\tan[\Delta\phi^\prime(r)]\,k(r)\,\sin[\alpha(r)], 
\end{equation} 
are typically as small compared to the radial steps of $\Delta\phi^\prime$,
i.e.  $|\der r|<70$~pc for all $r$ for NGC~5194.  In \S~\ref{sec:azcc}\ we
calculated $\Delta\phi^\prime$ through Eq.~\ref{eq:ccorr}, hence not only
within the spiral arms, but also as intensity-weighted mean across all the
azimuthal values.  If we were to calculate the angle-averaged value
$\Delta\phi^\prime$ from the average $\langle v_R\rangle$ and $\langle
v_\phi\rangle$ weighted by the product \hi~$\times$~24$\;\mu$m, which is the
weighting function of the cross-correlation, we find $\alpha\simeq 0$ for all
radii and a correction $k(r)$ even closer to unity than over a region limited
to the arms. With this approach we obtain $0.95<k(r)<1.05$ for NGC~5194.

\begin{table*}
\begin{center}
\caption{Characteristic Timescales \tHI\ and Pattern Speed $\Omega_p$
  Resulting from a $\chi^2$ Fit to the Angular Offsets $\Delta\phi$ After
  Correction for Non-circular Motions Following the Prescriptions in
  Section~\ref{sect:noncirc}}\label{tab:corromet} 
\begin{tabular}{lccc}
\tableline
 Obj. Name &  Alt. Name & \tHI  & $\Omega_p$ \\
 & &  [Myr] &  [km s$^{-1}$ kpc$^{-1}$] \\
\tableline
NGC 5194 & M51  & $3.3\pm0.6$ & $20\pm3$ \\
NGC 628  & M74  & $1.4\pm0.5$ & $26\pm3$ \\
NGC 6946 &      & $1.1\pm0.3$ & $36\pm4$ \\
\tableline
\end{tabular}
 \tablecomments{The best fits and the corrections are plotted in
  Fig.~\ref{fig:correction}. In comparison with the values listed in
  Table~\ref{tab:omet}, \tHI\ and $\Omega_p$ differ by less than their
  corresponding error bars.}
\end{center}
\end{table*}

After exploring the two extreme cases, (1) simple model of circular orbits and
(2) streaming motions near spiral arms for galaxies with prominent spiral
structure---where these effect are supposed to be the largest---we find that
the implied timescales \tHI\ do not vary significantly.  By estimating the
streaming motions in three galaxies from our data set, we find that the
correction $k(r)$ which we must apply to the angular offset measurements in
the scheme of circular orbits is generally near unity. Non-circular motions do
not greatly affect the offset measurements $\Delta\phi$ for the galaxies with
the most prominent spiral arms of our data set, where we expect indeed the
highest deviations from circular orbits, suggesting that, since
\tHI$\;\propto\Delta\phi$, the timescales \tHI\ will not vary by more than a
factor of 1.5.

However, these conclusions should be viewed cautiously, since (1) we are
assuming that $v_R$ and $v_\phi$ can be obtained by fitting
Eq.~\ref{eq:generalv}, and (2) we do not account for beam effects.  While the
presence of non-circular motions produces apparent radial variations of
inclination and position angle, our estimates are suggesting that these
effects do not influence much the determination of the timescales. Yet, tidal
interactions could ensue physical variations of inclination and position
angle, thus of the actual geometry of the observed velocities. Although we
note that among the sample galaxies only NGC~3031 (M81), NGC~3627, NGC~5055,
and NGC~5194 are affected by tidal interaction.  Moreover, we do not consider
extra-planar motions due to the implied numerical difficulties. For instance,
to include a vertical velocity component $v_z$ into Eq.~\ref{eq:generalv}\ 
would introduce a degeneracy while fitting $v_R$, $v_\phi$, and $v_z$, and a
degeneracy in the geometry of the motions, rendering the estimates of the
timescales uncertain.  However, after subtracting a circular orbit model from
the observed velocity field of the galaxy NGC~3184, an inspection of the
velocity residuals reveals deviations from circular motions of
$\sim$5--10~\kms\ amplitude, and about zero near the spiral arms.  Considering
that NGC~3184 is nearly face-on ($i=16$\dg), then vertical motions should not
exceed 5--10~\kms.  The beam deconvolution, on the other hand, would enlarge
the uncertainties on separating $v_R$ and $v_\phi$, but we rely on a
resolution size limit which is far below the typical thickness of a spiral arm
and allows us to resolve fine sub-structures within the arms. We also rely on
a large number of data points---several hundred to several thousand depending
on the galactocentric radius.

\section{DISCUSSION}\label{sec:discussion}

By analyzing the angular offsets between \hi\ and $24\;\mu$m in the context of
a simple kinematic model, we found short timescales, \tHI, as summarized in
the histogram of Figure~\ref{fig:hist}\ and in Table~\ref{tab:omet}. The
implied characteristic timescales for almost all sample galaxies lie in the
range 1--4 Myr.  This result sets an upper limit to the timescale for massive
star formation under these circumstances, since we are observing the time lag
between two phases: $a)$ the atomic gas phase, which subsequently is
compressed into molecular clouds and forms clusters of young, embedded,
massive stars, and $b)$ the warm dust phase, produced by heating from young
stars, whose UV radiation is reprocessed by the dust into the mid-IR, as
observed at 24$\;\mu$m. For the few objects where there are suitable CO data,
we checked that this geometric picture also holds for the molecular phase.

\subsection{Timescales Derived from Pattern Offsets}

\citet*{egusa04}\ used the same angular offset technique to compare CO and
\halpha\ emission. They report timescales \tCO~$\simeq4.8$~Myr for the galaxy
NGC~4254 (not in our sample). The \halpha\ traces a later evolutionary stage
than the warm dust emission, which can indicate the presence of a young
cluster still enshrouded by dust. Therefore, the \halpha\ and the dust
emission are expected to be separated by the time needed to remove the dusty
envelope, though they are observed to be spatially well correlated
\citep*{wong02,kennicutt98}. \citet*{prescott07}\ found a strong association
between 24~\mum\ sources and optical \hii\ regions in SINGS galaxies. Also
infrared sources located on top of older, UV-bright, clusters that do not have
\halpha\ emission are rare. Since \citet*{prescott07} suggest that the break
out time from dust clouds is short ($\sim$1~Myr), we do not expect a strong
offset.  \citet*{egusa04} derived the angular offset by subjective assessment
of the separation of the intensity peaks. They report that this may be a
source of systematic errors, since they cannot detect by eye, the angular
phase differences less than a certain threshold and in turn, their results
would possibly be an upper limit.  Given these considerations, the timescales
derived in this paper are likely consistent with the conclusions of
\citet*{egusa04}.

\citet*{rots75}\ and \citet*{garcia93}\ found time lags of $\sim$10~Myr for
M81 and M51, respectively. In particular, Rots applied the angular offset
method to the dust lanes and \halpha. This may be in part problematic since
the dust absorption in the optical bands only traces the presence of dust and
it is unrelated to the warm dust emission due to star-formation onset.
Garcia-Burillo et al. measured the spatial separation projected on the sky
between CO and \halpha\, and not the azimuthal offset.

\citet*{allen02}\ has argued that \hi\ is a photodissociation product of UV
shining on molecular gas, so it should be seen between the CO and UV/\halpha\ 
regions.  \citet*{allen86} observed \hi\ between spiral arm dust lanes and
\hii\ regions. However, \citet*{elmegreen07} points out that there is no time
delay between dust lanes and star formation: dust lanes may only represent a
heavy visual extinction effect and may not be connected to star-formation
onset. Our finding that CO is situated between \hi\ and hot dust (i.e.
Fig.~\ref{fig:HICO}) stands in conflict with the predictions of the model
proposed by \citet*{allen02}.

The evolutionary timescales of the ISM phases, especially for star formation,
are not well constrained.  The observational results of the last few decades,
arrive at different and, in some cases, controversial conclusions. The
discrepancies might in part be attributed to effects of limited resolution
(see \S~\ref{sec:spiralwave}).  This suggests that higher resolution and
sensitivity maps, especially for the CO emission, are needed to improve the
presented technique in the future.

\subsubsection{Photodissociation of H$_2$}

We argue that illumination effects of UV radiation shining on molecular and
dusty regions cannot photodissociate molecules in order to produce the
observed peaks of \hi---which correspond to typical surface densities of
several solar masses per pc$^2$. First, the mean free path of the UV photons
is remarkably short, typically $\sim$100~pc. The presence of dust,
particularly abundant in disk galaxies, is the main source of extinction in
particular within spiral arms, where we observe the peak of dust emission.
Second, none of the galaxies from our data set presents prominent nuclear
activity, whose UV flux could ionize preferentially the inner surfaces of the
molecular clouds. Also, we exclude that UV radiation from young stellar
concentrations can ionize preferentially one side of the clouds causing \hi\ 
and CO emissions to lie offset with respect to each other, which we instead
interpret as due to an evolutionary sequence. In fact, if the light from young
stars effectively produces \hi\ by photodissociation of H$_2$, then the
neutral to molecular gas fraction is expected to increase with star-formation
rate per unit area, $\Sigma_{\rm SFR}$, as a consequence of the increasing UV
radiation flux. Yet, the \hi\ to H$_2$ ratio decreases with increasing
$\Sigma_{\rm SFR}$, as also shown, for example, in \citet{kennicutt07}\ for
the galaxy M51. Moreover, the \hi\ density does not vary much as a function of
$\Sigma_{\rm SFR}$ \citep{kennicutt98}.

\subsection{Pattern Speeds}\label{sec:pspeeds}

The results of the fits in Figure~\ref{fig:dlr}\ are consistent with the
existence of a kinematic pattern speed for the considered galaxies, and are
suggesting that the spiral pattern must be metastable at least over a few Myr
or, strictly speaking, quasi-stationary. If the apparent spiral structure seen
in young stars were produced by stochastic self-propagating star formation
\citep[e.g.][]{Gerola1978,Seiden1982}\ and shear, without an underlying
coherent mass perturbation, then presumably we would not observe the
systematic radial variation of the offsets, seen in Figure~\ref{fig:dlr}, in
particular not that the offset changes sign at $R\simeq2-3\,R_{\rm exp}$. Our
results, however, do not exclude the stochastic star-formation mechanism to
occur, rather that this is not the dominant trigger of star formation.  The
spiral pattern might be the manifestation of full wealth of modes of
propagating density waves, which are continuously forming and dissolving gas
clouds and structures, where, i.e.  the azimuthal modes for a grand-design
spiral are dominant at low orders (e.g., $m=2,3,4$).  If the spiral structure
is quasi-stationary, then it can be characterized by an instantaneous pattern
speed to first order---at least over a timescale much shorter than the orbital
time.  This is opposed to density waves dynamically driven by bars or
interaction with companions, which could last a few orbital times.  Note,
however, that our analysis holds approximately in both possible cases.

Some of the fits in Figure~\ref{fig:dlr}\ do not accurately mirror the trend
of the observed data points, where the scatter is so large that it renders the
interpretation problematic. In particular, we could designate as a bad fit the
results for the galaxies NGC~3031, NGC~3621, NGC~7793, and NGC~6946
(Table~\ref{tab:omet}).  We also note that the pattern speed obtained in this
paper for the galaxies NGC~5055 and NGC~5194 disagree with previous
independent measurements \citep[e.g.][]{thornley1997,zimmer04}.  The fit for
NGC~5055 presents large error bars due to the large scatter in the azimuthal
offsets (see Figure~\ref{fig:dlr}), which may explain the differences.  For
NGC~5194 the large difference between our result and the Tremaine--Weinberg
method prediction could be due, as mentioned in \S~\ref{sec:corotation}, to
the assumption of continuity for the gas. In fact, \citet{zimmer04}\ find a
pattern speed $\Omega_p^Z\simeq 38$~\kms~kpc$^{-1}$ that is much faster than
the pattern speed predicted by the hydrodynamical models from \citet{kranz03},
$\Omega_p^K\simeq 12$~\kms~kpc$^{-1}$, using $R_{\rm cor}\simeq 2.7\,R_s$ and
$R_s\simeq1.4'$.  Moreover, the presence of large variations of the azimuthal
offsets as a function of radius with respect to the smoother fitted curves of
Figure~\ref{fig:dlr}\ suggests that the pattern speed may not be constant over
the entire disk. Instead, the spiral pattern could be described by more than
one pattern speed, implying that the delay time is not exactly the same in all
parts of an individual galaxy, and not necessarily the same in all the
considered galaxies. However, the offsets measurements and the implied pattern
speeds and timescales could be statistically pointing toward regions of high
\hi\ and 24~\mum\ fluxes when using the weighting of Equation~\ref{eq:ccorr},
therefore toward high $\Sigma_{\rm SFR}$, ensuing the timescales to be
comparable in all cases.

\subsection{Is Star Formation Triggered by Spiral Waves?}\label{sec:spiralwave}

Addressing to our results (\S~\ref{sec:results} and Fig.~\ref{fig:dlr}), a
further aspect emerging from our analysis concerns the same general behavior
displayed by an heterogeneous sample of galaxies going from grand-design (e.g.
NGC~5194) to flocculent morphologies (e.g. NGC~2841 and NGC~5055, see also
Fig.~\ref{fig:reprex}\ for an example comparison).  Visual examination of the
IR band images for our sample galaxies (e.g. at 3.6~\mum) indicates that
almost all have two-arm or multi-arm coherent spiral arms; for none of these
galaxies the spiral structure is so chaotic as to be characterized as
flocculent.  Moreover, both types of galaxies from our sample display
comparable integrated star-formation rates.  Grand-design density waves are
not likely to be the primary trigger of star formation, since a substantial
portion of stars are also formed in the inter-arm regions \citep{elmegreen86}.
The concentration of young stellar populations near prominent spiral arms is
rather an effect of kinematics \citep{roberts69}.  Grand-design and flocculent
galaxies exhibit the same intrinsic self-similar geometry
\citep*{elmegreen03}.  The difference between these two types of galaxies is
only dictated by a different distribution of azimuthal modes. Large scales
structures (low-order modes) are the dominant features in grand design.
However, the short timescales suggest that the physical scales where star
formation occurs are rather small, since this time delay cannot exceed the
free-fall time.  Weak density waves, those described by high-order azimuthal
modes, not necessarily grand-design modes, are likely to facilitate the growth
of super-critical structures which end in star-forming events.  In galaxies
morphologically classified as flocculent and grand design, the mechanism that
triggers star formation must be the same---gravitational instability.  Direct
compression of gas clouds is indeed able to locally trigger star formation.

The sizes of the structures of active star formation, however, are not
representative of the timescales involved.  Star-forming regions are organized
hierarchically according to the small-scale turbulent motions of gas and
stars, where the dynamical time varies as a function of the local scale sizes
\citep{efremov98,ballesteros99}. The improved accuracy of recent observational
techniques, e.g. THINGS and SINGS, allows us to observe smaller and smaller
structures, which evolve thus more rapidly and are characterized by shorter
timescales.  Stellar activity and turbulence limit the lifetime of molecular
clouds by causing the destruction of the parent cloud and the cloud dispersal,
respectively. This lifetime, typically $10-20$~Myr, drops by a factor of
$\sim$10 for the star-forming clouds.  Moreover, star formation begins at high
rate in only a few Myr, appearing in large structures of O-B complexes---beads
on a string---of few hundred pc scales, and remaining active for
$\sim$30--50~Myr, but with a gradually decreasing star-forming rate, until the
complete quenching \citep{elmegreen07}.  The timescales measured here refer to
the very initial phase of star formation, specifically when star-formation
rate is the highest---traced by 24~\mum\ peak emission.  The time delay \tHI\ 
needs not to represent the average time difference between these two phases,
especially if there is a gradually declining tail of star-formation activity,
taking longer than \tHI. If much of the star formation were actually occurring
within our short \tHI\ estimate, then star formation would have to be
inefficient to avoid conflicts with the short-term depletion of the gas
reservoir.

\subsection{Can We Rule Out Timescales of 10 Myr?}

Since the 24$\;\mu$m emission traces the mass-weighted star-formation
activity, the timescale \tHI\ measures the time for star clusters to form from
\hi\ gas, but it does not show that all molecular clouds live only $\sim
$2~Myr. It might show that the bulk of massive stars that form in disks has
emerged from molecular clouds that only lived $\sim$2~Myr, though there could
still be molecular clouds that live an order of magnitude longer.  The full
molecular cloud lifetimes estimated for the LMC correspond to $\sim$10~Myr
\citep{Mizuno2001,Yamaguchi2001}, while star clusters are formed from
molecular clouds in only few Myr.  Yet, more recent observations suggest much
slower evolution. In particular, \citet{blitz06}\ propose molecular cloud
lifetimes for the LMC as long as 20--30~Myr. In the Milky Way this timescale
is estimated to be a few Myr \citep*[e.g. ][]{HBB01}, but the examined cloud
complexes, such as Taurus and Ophiuchus, have low star-formation rate and
masses more than an order of magnitude lower than those studied by
\citet*{blitz06}.  Thus, it remains possible that our results and these
previous arguments are all consistent.  At high density star formation also
occurs at higher rate. \citet*{blitz06} also find that the timescales for the
emergence of the first \hii\ regions traced by \halpha\ in molecular clouds is
$t_{\rm HII}\sim 7\;$Myr, which does not exclude that $t_{\rm HII}\ge t_{{\rm
    HI}\mapsto 24\,\mu{\rm m}}$, but it could be problematic with the short
break out suggested by \citet*{prescott07}, since the time delay which is
required between the onset of star formation (as traced by 24$\;\mu$m from
obscured \hii\ regions) and the emergence of \halpha\ emission would need to
be large.

In conclusion, we note that the timescale \tHI\ results from a $\chi^2$ fit to
all the data, and it is treated as a global constant individually for each
galaxy, so it is not a function of radius. Globally, all the characteristic
timescales \tHI\ listed in Table~\ref{tab:omet}\ are $\leq$4~Myr, except one
single case (NGC~925). The error bars are also relatively small: $<$1~Myr for
the majority of the cases.  These results clearly exclude characteristic
timescales \tHI\ of the order of $\sim$10~Myr, even for the highest value
recorded in our data set which is NGC~925.

\subsection{Theoretical Implications}

The short timescale found here between the peak of \hi\ emission and the peak
of emission from young, dust-enshrouded stars has implications for two related
theoretical controversies.  First is the question of whether molecular clouds
are short-lived, dynamically evolving objects (\citealt*{ballesteros99},
\citealt{HBB01,elmegreen2000, ballesteros06, elmegreen07}) or quasi-static
objects evolving over many free-fall times
\citep*{matzner02,KrumholzMatzner06}.  The second, related question is what
the rate-limiting step for star formation in galaxies is: formation of
gravitationally unstable regions in the \hi\ that can collapse into molecular
clouds \citep{elmegreen02, kravtsov03, li05, li06, elmegreen07}, or formation
of dense, gravitationally unstable cores within quasi-stable molecular clouds
\citep*{krumholz05,KrumholzMatzner06,KrumholzTan07}.

The short timescales found here for the bulk of massive star formation in
regions of strong gravitational instability appears to support the concept
that molecular cloud evolution occurs on a dynamical time once gravitational
instability has set in, and that the rate-limiting step for star formation is
the assembly of \hi\ gas into gravitationally unstable configurations.  Our
work does not, however, address the total lifetime of molecular gas in these
regions, as we only report the separation between the peaks of the emission
distributions.  Molecular clouds may well undergo an initial burst of star
formation that then disperses fragments of molecular gas that continues star
formation at low efficiency for substantial additional time
\citep*{elmegreen07}.  Averaging over the efficient and inefficient phases of
their evolution might give the overall low average values observed in galaxies
\citep*[e.g.][]{KrumholzTan07}.

\section{CONCLUSIONS}\label{sec:conclusions}

We have derived characteristic star-formation timescales for a set of nearby
spiral galaxies, using a simple geometric approach based on the classic
\citet{roberts69}\ picture that star formation occurs just downstream from the
spiral pattern, where gas clouds have been assembled into super-critical
configurations. This derived timescale, \tHI, refers to the processes from the
densest \hi, to the molecular phase, to enshrouded hot stars heating the dust.
The analysis is based on high-resolution 21~cm maps from THINGS, which we
combined with $24\;\mu$m maps from SINGS. We assume that the observed spiral
arms have a pattern speed $\Omega_p$. Given the rotation curve,
$v_c(r)=r\,\Omega(r)$, this allows us to translate angular offsets at
different radii between the \hi\ flux peaks and the $24\;\mu$m flux peaks in
terms of a characteristic time difference.

At each individual point along the spiral arm we found considerable scatter
between the \hi\ and $24\;\mu$m emission peaks. However, for each galaxy we
could arrive at a global fit, using a cross-correlation technique, and derive
two characteristic parameters, \tHI\ and $R_{\rm cor}$. For our 14 objects we
found the general relation $R_{\rm cor}=(2.7\pm0.2)\,R_s$, which is consistent
with previous studies \citep*[e.g.][]{kranz03}\ and, more importantly, we
found \tHI\ to range between 1 and 4~Myr. Even when accounting for
uncertainties, at the highest peak of star-formation rate timescales as long
as \tHI$\;\sim10\,$Myr, which have been inferred from other approaches, do not
appear consistent with our findings. At least for the case of nearby spiral
galaxies, our analysis sets an upper limit to the time needed to form massive
stars (responsible for heating the dust) by compressing the (atomic) gas.
Therefore, it points to a rapid procession of star formation through the
molecular-cloud phase in spiral galaxies. If star formation really is as rapid
as our estimate of \tHI\ suggests, it must be relatively inefficient to avoid
the short-term depletion of gas reservoirs.

\acknowledgments We thank Henrik Beuther, Mark Krumholz, Adam Leroy, and Eve
Ostriker for useful discussions and suggestions. We are also grateful to the
anonymous referee, whose comments helped us to improve the manuscript. The
work of W.J.G.d.B. is based upon research supported by the South African
Research Chairs Initiative of the Department of Science and Technology and
National Research Foundation. E.B. gratefully acknowledges financial support
through an EU Marie Curie International Reintegration Grant (Contract No.
MIRG-CT-6-2005-013556).  M.-M.M.L. was partly supported by US National Science
Foundation grant AST 03-07854, and by stipends from the Max Planck Society and
the Deutscher Akademischer Austausch Dienst. This research has made use of the
NASA/IPAC Extragalactic Database (NED) which is operated by the Jet Propulsion
Laboratory, California Institute of Technology, under contract with the
National Aeronautics and Space Administration.

\end{document}